\documentclass[10pt]{article}
\usepackage{amsbsy}
\newcommand{\beq}{\begin{equation}}
\newcommand{\eeq}{\end{equation}}
\newcommand{\bdm}{\begin{displaymath}}
\newcommand{\edm}{\end{displaymath}}
\newcommand{\beqr}{\begin{eqnarray}}
\newcommand{\eeqr}{\end{eqnarray}}
\newcommand{\beqrn}{\begin{eqnarray*}}
\newcommand{\eeqrn}{\end{eqnarray*}}
\def\l{\lambda}

\def\R{{\bf R}}

\def\bchi{\boldsymbol{\chi}}

\def\ve{\varepsilon}
\def\a{\alpha}
\def\k{\kappa}

\begin{document}

\title{Explicit computations of low lying eigenfunctions for the quantum trigonometric Calogero-Sutherland model related to the exceptional algebra $E_7$ }

\author{J. Fern\'andez N\'u\~{n}ez, W. Garc\'{\i}a Fuertes,  A.M. 
Perelomov\footnote{On leave of absence from the Institute for Theoretical and Experimental Physics, 117259, Moscow, Russia. Current E-mail address: perelomo@dftuz.unizar.es}\\     \normalsize {\it Departamento de F\'{\i}sica, Facultad de Ciencias, Universidad de Oviedo, E-33007 Oviedo, Spain}}
\date{}

\maketitle

\begin{abstract}
In a previous paper \cite{fgp05c} we have studied the characters and Clebsch-Gordan series for the exceptional Lie algebra $E_7$ by relating them to the quantum trigonometric Calogero-Sutherland Hamiltonian with coupling constant $\kappa=1$. Now we extend that approach to the case of general $\kappa$.
\end{abstract}

\section{Introduction}
The Calogero-Sutherland models \cite{ca71,su72} related to the root systems of the simple Lie algebras \cite{op77,op78,op83} have been deeply investigated during the last two decades. Originally introduced on purely theoretical grounds, this class of models have however found a number of relevant applications in such diverse fields as condensed matter physics, supersymmetric Yang-Mills theory or black-hole physics. On the mathematical side, an interesting feature of the quantum version of this kind of models is that their energy eigenfunctions provide a natural generalization of several types of hypergeometric functions to the multivariable case. For the potential $v(q)=\kappa(\kappa-1)\sin^{-2}(q)$ and special values of the coupling constant, these eigenfunctions are related to some othogonal functional systems of particular interest in the theory of Lie algebras and symmetric spaces: for $\kappa=1$ we obtain the characters of the irreducible representations of the algebra, while for $\kappa=0$ the corresponding monomial symmetric functions arise; other values of $\kappa$  lead to zonal spherical functions in symmetric spaces associated to the Lie algebra: in particular, for $E_7$, $\kappa=\frac{1}{2}$ gives these functions for the symmetric space E$\,$V$^*$ \cite{he78,op83}. The Calogero-Sutherland Hamiltonian appears in this way as a natural unified tool for the computation of all these objects.

The Calogero-Sutherland Hamiltonian associated to the root system of a simple Lie algebra can be written as a second-order differential operator whose variables are the characters of the fundamental representations of the algebra.  As it was shown in the papers \cite{pe98a,prz98,pe99}, and later in \cite{pe00,flp01,fp02,ffp03,ngp03, ngp051, ngp05b}, this approach gives the possibility of developing some systematic procedures to solve the Schr\"{o}dinger equation and determine important properties of the eigenfunctions, such as recurrence relations or generating functions for some subsets of them. The approach has been used for classical algebras of $A_n$ and $D_n$ type, for the exceptional algebra $E_6$, and recently also for $E_7$ for the special value of the coupling constant for which the eigenfunctions are proportional to the characters of the irreducible representations of the algebra. The aim of this paper is to show how to generalize the treatment given in \cite{fgp05c} to arbitrary values of the coupling constant and to extend some of the particular results found there to the general case.
\section{The Calogero-Sutherland Hamiltonian for $E_7$ in Weyl-invariant variables}
The  trigonometric Calogero-Sutherland model related to the root system $\cal R$ of a simply-laced Lie algebra of rank $r$ is the quantum system in an Euclidean space $\R^r$ defined by the standard Hamiltonian operator 
\beq
\label{ham}
H=\frac{1}{2}\sum_{j=1}^rp_j ^2+\sum_{\a\in{\cal R}^+}\kappa_\a(\kappa_\a-1)\sin^{-2}(\a,q),
\eeq
where $q=({q_j})$ is  a cartesian coordinate system and $p_j=-{\rm i}\,\partial_{q_j}$; ${\cal R}^+$  is the set of the positive roots of the algebra, and $\kappa$ is the coupling constant. The (non-normalized) ground state wave function is
\beq
\Psi_0^\k(q)={\prod_{\a\in {\cal R}^+}\sin^\k(\a, q)},
\eeq
while the excited states are indexed by the highest weights $\mu=\sum m_i\l_i\in P^+$ ($P^+$ is the cone of dominant weights) of the irreducible representations of the algebra, that is, by the $r$-tuple of non-negative integers ${\bf m}=(m_1,\dots,m_r)$. Looking for solutions $\Psi_{\bf m}^\k$ of the Schr\"{o}dinger equation  in the form 
\beq
\Psi_{\bf m}^\k(q)=\Psi_0^\k(q)\Phi_{\bf m}^\k(q),
\eeq
we are led to the eigenvalue problem
\beq
\Delta^\k\Phi_{\bf m}^\k=\ve_{\bf m}(\k)\Phi_{\bf m}^\k\,,
\label{sch}
\eeq
where $\Delta ^\k$ is the linear differential operator
\beq
\Delta^\k=-\frac{1}{2}\sum_{j=1}^r\partial_{q_j}^{\,2}-\k\sum_{\a\in {\cal R}^+}  {\cot}(\a, q)(\a,\nabla_q)
\label{4b}.
\eeq
Due to the Weyl symmetry of the Hamiltonian, to solve the eigenvalue problem (\ref{sch}) it is convenient to express the operator $\Delta^\kappa$ in a set of independent $W$-invariant variables such as $z_k=\bchi_{\l_k}(q)$, the characters of the irreducible representations of the algebra. The operator $\Delta^\k$ in the $z$-variables has the structure:
\beq
\Delta^\kappa=\sum_{j, k} a_{jk}(z)\partial_{z_j}\partial_{z_k}+\sum_{j} \left[b_j^{0}(z)+\kappa\, b_j^{1}(z)\right]\partial_{z_j},
\label{deltaz}
\eeq
but to fix the coefficients by direct change of variables is very cumbersome. As explained in \cite{fgp05c}, a different procedure, based on the computation of the quadratic Clebsch-Gordan series and the second order characters of $E_7$, is possible. In \cite{fgp05c}, we applied this procedure  to compute $a_{jk}(z)$ and $b_j^0(z)+b_j^1(z)$, thus finding the operator $\Delta^\kappa$ for the special case $\kappa=1$. The remaining task is to compute $b_j^0(z)$ and $b_j^1(z)$ separately.

To accomplish that task a new piece of information is required: we need to know all the first order symmetric monomials for $E_7$ given as a function of the $z$-variables. To obtain them, we will rely on the expansions of the fundamental characters of $E_7$ in terms of monomial functions computed by Loris and Sasaki in \cite{ls04}. In the notation of \cite{fgp05c}, these expansions are
\begin{eqnarray*}
z_1&=&M_{\lambda_1}+7,\\
z_2&=&M_{\lambda_2}+6 M_{\lambda_7},\\
z_3&=&M_{\lambda_3}+5 M_{\lambda_6}+22 M_{\lambda_1}+77,\\
z_4&=&M_{\lambda_4}+4 M_{\lambda_1+\lambda_6}+15 M_{\lambda_2+\lambda_7}+ 15 M_{2\lambda_1}+45 M_{2\lambda_7}+50 M_{\lambda_3}+145 M_{\lambda_6}+390 M_{\lambda_1}+980,\\
z_5&=&M_{\lambda_5}+5 M_{\lambda_1+\lambda_7}+21 M_{\lambda_2}+71 M_{\lambda_7},\\
z_6&=&M_{\lambda_6}+6 M_{\lambda_1}+27,\\
z_7&=&M_{\lambda_7}.
\end{eqnarray*}
To invert these formulas to compute the fundamental monomial functions, we have to proceed in increasing order of the height of the dominant weights associated to the characters. Once a first-order monomial function is known, we compute the corresponding $b_k^0(z)$ and $b_k^1(z)$ following the procedure described in \cite{fgp05c}. This makes it possible to use the part of the operator $\Delta^\kappa$ already known in each step to compute the second order monomial functions in advance, i.e. before they are needed to obtain the next fundamental monomial function. With this strategy, it is easy to find
\begin{eqnarray*}
M_{\lambda_1}&=&z_1-7,\\
M_{\lambda_2}&=&z_2-6z_7,\\
M_{\lambda_3}&=&z_3-5 z_6+ 8 z_1+2,\\
M_{\lambda_4}&=&z_4-4 z_1 z_6+9 z_2 z_7+ 9 z_1^2+9 z_7^2-14 z_3-39 z_6-22 z_1-18,\\
M_{\lambda_5}&=&z_5-5 z_1 z_7+14 z_2+15 z_7,\\
M_{\lambda_6}&=&z_6-6 z_1+15,\\
M_{\lambda_7}&=&z_7,
\end{eqnarray*}
and, therefore,
\begin{eqnarray*}
b_1^0(z)+\kappa b_1^1(z)&=&-28 + 4 z_1 + \kappa (28 + 68 z_1)\\
b_2^0(z)+\kappa b_2^1(z)&=&7 z_2 - 24 z_7 + \kappa (98 z_2 + 24 z_7)\\
b_3^0(z)+\kappa b_3^1(z)&=&8 - 56 z_1 + 12 z_3 - 20 z_6 + 
  \kappa (-8 + 56 z_1 + 132 z_3 + 20 z_6)\\
b_4^0(z)+\kappa b_4^1(z)&=&-72 + 72 z_1 - 24 z_1^2 + 24 z_3 + 
  24 z_4 - 16 z_6 - 16 z_1 z_6 - 
  24 z_2 z_7 + 36 z_7^2 + \\
 & & \kappa (72 - 72 z_1 + 24 z_1^2 - 24 z_3 + 
     192 z_4 + 16 z_6 + 16 z_1 z_6 + 
     24 z_2 z_7 - 36 z_7^2)\\
b_5^0(z)+\kappa b_5^1(z)&=&-28 z_2 + 15 z_5 - 4 z_7 - 20 z_1 z_7 + 
  \kappa (28 z_2 + 150 z_5 + 4 z_7 + 
     20 z_1 z_7)\\
b_6^0(z)+\kappa b_6^1(z)&=&-48 - 24 z_1 + 8 z_6 + 
  \kappa (48 + 24 z_1 + 104 z_6)\\
b_7^0(z)+\kappa b_7^1(z)&=&3 z_7 + 54 \kappa z_7.
\end{eqnarray*}

This completes the computation of $\Delta^\kappa$. In the rest of the paper we present some results obtained through the use of this operator.

\section{Some explicit results on the low lying eigenfunctions of the systems}

In this Section, we present some results on the first and second order polynomials and generalized quadratic Clebsch-Gordan series. Because some formulas are too long, we give the complete results, in a form suitable for use in Mathematica or Maple, in the adjoint files {\tt results31.txt} -- {\tt results35.txt}, which are accessible through the ``source" format of this document.

\subsection{Second order monomial symmetric functions}
Once we know $\Delta^\kappa$, we can compute its eigenfunctions by means of the iterative algorithms given in \cite{fgp05c}. In particular, we can obtain the monomial symmetric functions for $E_7$ by simply taken $\kappa=0$ in these algorithms. We present here the list of the second order monomial functions obtained in that way.
\begin{eqnarray*}
M_{2000000}&=&z_1^2
-2 z_3
-2 z_1
-7
\\
M_{1100000}&=&z_1 z_2
-5 z_5
+3 z_1 z_7
-23 z_7
\\
M_{0200000}&=&z_2^2
-2 z_4
-2 z_2 z_7
-2 z_1^2
-6 z_7^2
+4 z_3
+14 z_6
+4 z_1
+12
\\
M_{1010000}&=&z_1 z_3
-3 z_4
-z_1 z_6
+6 z_2 z_7
-3 z_1^2
-9 z_7^2
-9 z_3
+4 z_6
+20 z_1
+32
\\
M_{0110000}&=&z_2 z_3
-4 z_1 z_5
+5 z_2 z_6
+4 z_1^2 z_7
-4 z_3 z_7
-17 z_1 z_2
-16 z_6 z_7
+41 z_5
+13 z_1 z_7
+12 z_2
+7 z_7
\\
M_{0020000}&=&z_3^2
-2 z_1 z_4
+2 z_2 z_5
-2 z_3 z_6
-7 z_6^2
+12 z_5 z_7
-2 z_1^3
+4 z_1 z_3
+2 z_1 z_7^2
-10 z_4
-2 z_1 z_6
+10 z_1^2\\
&+&10 z_7^2
-16 z_3
-22 z_6
-16 z_1
-8
\\
M_{1001000}&=&z_1 z_4
-4 z_2 z_5
-4 z_1^2 z_6
+10 z_3 z_6
+9 z_1 z_2 z_7
+14 z_6^2
-34 z_5 z_7
+9 z_1^3
-21 z_2^2
-39 z_1 z_3
-21 z_1 z_7^2
\\
&+&66 z_4
+54 z_1 z_6
+23 z_2 z_7
+36 z_1^2
-22 z_7^2
-54 z_3
-24 z_6
-56 z_1
-24
\\
M_{0101000}&=&z_2 z_4
-3 z_3 z_5
+2 z_1 z_2 z_6
-2 z_2^2 z_7
+5 z_1 z_3 z_7
+5 z_5 z_6
-14 z_4 z_7
-19 z_1 z_6 z_7
-12 z_1^2 z_2
+15 z_2 z_3
\\
&+&17 z_2 z_7^2
+25 z_1 z_5
+19 z_1^2 z_7
+42 z_7^3
+5 z_2 z_6
-29 z_3 z_7
-27 z_1 z_2
-133 z_6 z_7
+131 z_5
+10 z_1 z_7
\\
&+&40 z_2
-43 z_7
\\
M_{0011000}&=&z_3 z_4
-3 z_1 z_2 z_5
+5 z_2^2 z_6
+2 z_1 z_3 z_6
+5 z_5^2
-7 z_4 z_6
+5 z_1^2 z_2 z_7
-10 z_1 z_6^2
-14 z_2 z_3 z_7
+10 z_1 z_5 z_7
\\
&-&12 z_1^2 z_3
-3 z_1^2 z_7^2
-5 z_1 z_2^2
+24 z_3^2
+5 z_1 z_4
-2 z_2 z_6 z_7
+6 z_1^2 z_6
+z_2 z_5
+11 z_3 z_7^2
+10 z_3 z_6
\\
&+&15 z_6 z_7^2
-28 z_1 z_2 z_7
-24 z_6^2
+40 z_2^2
+4 z_5 z_7
+21 z_1^3
-15 z_1 z_3
+19 z_1 z_7^2
-16 z_4
-54 z_1 z_6
\\
&-&48 z_1^2
+17 z_2 z_7
-5 z_7^2
+7 z_3
+31 z_6
-16 z_1
-22
\\
M_{0002000}&=&z_4^2
-2 z_2 z_3 z_5
+2 z_3^2 z_6
+2 z_1 z_5^2
+2 z_1 z_2^2 z_6
-2 z_2^3 z_7
-4 z_1 z_4 z_6
-2 z_1 z_2 z_3 z_7
-2 z_2 z_5 z_6
+6 z_2 z_4 z_7
\\
&-&8 z_1^2 z_6^2
+12 z_3 z_6^2
+12 z_1^2 z_5 z_7
-20 z_3 z_5 z_7
-6 z_1^2 z_4
+2 z_1 z_2 z_6 z_7
+2 z_6^3
+2 z_2^2 z_3
+2 z_1^3 z_7^2
\\
&-&4 z_5 z_6 z_7
+9 z_2^2 z_7^2
-4 z_1 z_3 z_7^2
+12 z_3 z_4
-4 z_1 z_2 z_5
-16 z_2^2 z_6
-14 z_4 z_7^2
-4 z_1^3 z_6
+2 z_5^2
+8 z_1 z_3 z_6
\\
&+&2 z_1^2 z_2 z_7
+30 z_4 z_6
+9 z_1^4
-18 z_2 z_7^3
+12 z_1 z_6^2
-6 z_2 z_3 z_7
-16 z_1 z_5 z_7
-36 z_1^2 z_3
+4 z_1 z_2^2
-8 z_1^2 z_7^2
\\
&+&26 z_3^2
+32 z_2 z_6 z_7
+16 z_3 z_7^2
+32 z_1 z_4
-28 z_2 z_5
+14 z_1^2 z_6
+9 z_7^4
-22 z_6 z_7^2
-16 z_1^3
-5 z_6^2
\\
&-&8 z_1 z_2 z_7
+32 z_1 z_3
-16 z_2^2
+52 z_5 z_7
-48 z_4
-10 z_1 z_7^2
+36 z_2 z_7
+42 z_7^2
+20 z_1^2
-8 z_3
-88 z_6
\\
&-&8 z_1
-60
\\
M_{1000100}&=&z_1 z_5
-5 z_2 z_6
-5 z_1^2 z_7
+15 z_3 z_7
+9 z_1 z_2
+19 z_6 z_7
-54 z_5
-29 z_1 z_7
+30 z_2
+56 z_7
\\
M_{0100100}&=&z_2 z_5
-4 z_3 z_6
+4 z_6^2
+5 z_1 z_2 z_7
-7 z_2^2
+4 z_1 z_3
-4 z_5 z_7
-10 z_4
-16 z_1 z_7^2
+4 z_1 z_6
+29 z_2 z_7
\\
&+&54 z_7^2
+6 z_1^2
-12 z_3
-90 z_6
-76 z_1
+4
\\
M_{0010100}&=&z_3 z_5
-4 z_1 z_2 z_6
+9 z_2^2 z_7
+5 z_1 z_3 z_7
+5 z_5 z_6
-12 z_4 z_7
-11 z_1 z_6 z_7
+4 z_1^2 z_2
-25 z_2 z_3
+16 z_1 z_5
\\
&-&4 z_1^2 z_7
-5 z_2 z_6
+19 z_3 z_7
-2 z_1 z_2
+31 z_6 z_7
-46 z_5
-26 z_1 z_7
-40 z_2
+74 z_7
\\
M_{0001100}&=&z_4 z_5
-3 z_2 z_3 z_6
+2 z_1 z_5 z_6
+5 z_3^2 z_7
+5 z_2 z_6^2
+5 z_1 z_2^2 z_7
-7 z_1 z_4 z_7
-7 z_2^3
-12 z_1 z_2 z_3
-14 z_2 z_5 z_7
\\
&+&27 z_2 z_4
-10 z_1^2 z_6 z_7
+10 z_3 z_6 z_7
-2 z_1 z_2 z_7^2
-3 z_6^2 z_7
+21 z_1^2 z_5
-17 z_3 z_5
+10 z_1 z_2 z_6
+11 z_5 z_7^2
\\
&+&5 z_1^3 z_7
-8 z_5 z_6
+15 z_1 z_7^3
+6 z_2^2 z_7
-5 z_1 z_3 z_7
-5 z_1^2 z_2
-17 z_4 z_7
+16 z_2 z_3
-26 z_1 z_6 z_7
-16 z_1 z_5
\\
&-&45 z_2 z_7^2
-z_1^2 z_7
+45 z_2 z_6
+5 z_7^3
+31 z_3 z_7
-26 z_1 z_2
+14 z_6 z_7
+42 z_5
-26 z_1 z_7
+110 z_2
-26 z_7
\\
M_{0000200}&=&z_5^2
-2 z_4 z_6
+2 z_2 z_3 z_7
-2 z_3^2
-2 z_1 z_5 z_7
-2 z_1 z_2^2
-7 z_1^2 z_7^2
+4 z_1 z_4
+12 z_3 z_7^2
+2 z_2 z_5
+14 z_1^2 z_6
\\
&-&24 z_3 z_6
+2 z_1 z_2 z_7
+2 z_6 z_7^2
+4 z_1^3
-4 z_6^2
+2 z_5 z_7
+14 z_2^2
+10 z_1 z_7^2
-8 z_1 z_3
-28 z_4
-24 z_1 z_6
\\
&-&26 z_2 z_7
+17 z_7^2
-20 z_1^2
+32 z_3
-8 z_6
+32 z_1
-32
\\
M_{1000010}&=&z_1 z_6
-6 z_2 z_7
+9 z_7^2
-6 z_1^2
+21 z_3
+6 z_6
-27 z_1
+27
\\
M_{0100010}&=&z_2 z_6
-5 z_3 z_7
+3 z_6 z_7
+9 z_1 z_2
-7 z_5
-19 z_1 z_7
+15 z_2
+7 z_7
\\
M_{0010010}&=&z_3 z_6
-5 z_6^2
-5 z_1 z_2 z_7
+14 z_2^2
+9 z_1 z_3
+15 z_5 z_7
-27 z_4
+19 z_1 z_7^2
-49 z_1 z_6
-14 z_2 z_7
-27 z_1^2
\\
&-&4 z_7^2
+42 z_3
+79 z_6
+12 z_1
-42
\\
M_{0001010}&=&z_4 z_6
-4 z_2 z_3 z_7
-4 z_1 z_6^2
+9 z_3^2
+10 z_1 z_5 z_7
+9 z_2 z_6 z_7
+9 z_1 z_2^2
+14 z_1^2 z_7^2
-18 z_1 z_4
-34 z_3 z_7^2
\\
&-&33 z_2 z_5
-39 z_1^2 z_6
+72 z_3 z_6
-5 z_1 z_2 z_7
-21 z_6 z_7^2
-18 z_1^3
+34 z_6^2
+22 z_5 z_7
-14 z_2^2
+18 z_1 z_7^2
\\
&+&48 z_1 z_3
-7 z_4
-16 z_2 z_7
-61 z_7^2
-27 z_1^2
+18 z_3
+151 z_6
+150 z_1
+26
\\
M_{0000110}&=&z_5 z_6
-3 z_4 z_7
-z_1 z_6 z_7
+5 z_2 z_3
+6 z_2 z_7^2
-2 z_1 z_5
-11 z_2 z_6
-9 z_7^3
-9 z_1^2 z_7
+9 z_3 z_7
+7 z_1 z_2
\\
&+&25 z_6 z_7
-16 z_5
+26 z_1 z_7
-60 z_2
+16 z_7
\\
M_{0000020}&=&z_6^2
-2 z_5 z_7
+2 z_4
-6 z_1^2
-2 z_7^2
+12 z_3
+4 z_6
+12 z_1
+17
\\
M_{1000001}&=&z_1 z_7
-7 z_2
+8 z_7
\\
M_{0100001}&=&z_2 z_7
-6 z_3
-6 z_7^2
+14 z_6
+16 z_1
-28
\\
M_{0010001}&=&z_3 z_7
-5 z_6 z_7
-6 z_1 z_2
+20 z_5
+24 z_1 z_7
-49 z_2
-12 z_7
\\
M_{0001001}&=&z_4 z_7
-4 z_1 z_6 z_7
-5 z_2 z_3
+9 z_2 z_7^2
+14 z_1 z_5
-5 z_2 z_6
+9 z_7^3
+19 z_1^2 z_7
-44 z_3 z_7
-14 z_1 z_2
\\
&-&53 z_6 z_7
+54 z_5
+42 z_1 z_7
-40 z_2
-144 z_7
\\
M_{0000101}&=&z_5 z_7
-4 z_4
-5 z_1 z_7^2
+8 z_1 z_6
+11 z_2 z_7
+12 z_1^2
-3 z_7^2
-42 z_3
-18 z_6
-4
\\
M_{0000011}&=&z_6 z_7
-3 z_5
-z_1 z_7
+7 z_2
-11 z_7
\\
M_{0000002}&=&z_7^2
-2 z_6
-2
\end{eqnarray*}
\subsection{Expansion of second order characters in monomial functions}
As the name suggests, the orthogonal system of monomial symmetric functions is the simplest one among the different classes of symmetric polynomials associated to the Lie algebra $E_7$: each monomial symmetric function is nothing but the sum of all the monomials associated to one orbit of the Weyl group on the weight lattice. Now, we can easily expand other polynomials associated to the root system of $E_7$ in the basis of the monomial symmetric functions. In fact, the method is the same which we have described in \cite{fgp05c} for the computation of Clebsch-Gordan series. In particular, the coefficients in the decomposition of characters in monomial symmetric functions are interesting in that they give the multiplicities of the weights in the corresponding irreducible representations. As an example, we present such decomposition for all the second order characters.
\begin{eqnarray*}
\bchi_{2 0 0 0 0 0 0}&=&
 M_{2 0 0 0 0 0 0}
+    M_{0 0 1 0 0 0 0}
+ 4 M_{0 0 0 0 0 1 0}
+ 17 M_{1 0 0 0 0 0 0}
+ 63 M_{0 0 0 0 0 0 0}
\\
\bchi_{1 1 0 0 0 0 0}&=&
 M_{1 1 0 0 0 0 0}
+ 4 M_{0 0 0 0 1 0 0}
+ 16 M_{1 0 0 0 0 0 1}
+ 56 M_{0 1 0 0 0 0 0}
+ 171 M_{0 0 0 0 0 0 1}
\\
\bchi_{0 2 0 0 0 0 0}&=&
 M_{0 2 0 0 0 0 0}
+    M_{0 0 0 1 0 0 0}
+ 3 M_{1 0 0 0 0 1 0}
+ 11 M_{0 1 0 0 0 0 1}
+ 10 M_{2 0 0 0 0 0 0}
+ 36 M_{0 0 0 0 0 0 2}
+ 34 M_{0 0 1 0 0 0 0}\\ 
&+& 96 M_{0 0 0 0 0 1 0}
+ 248 M_{1 0 0 0 0 0 0}
+ 603 M_{0 0 0 0 0 0 0}
\\
\bchi_{1 0 1 0 0 0 0}&=&
 M_{1 0 1 0 0 0 0}
+ 2 M_{0 0 0 1 0 0 0}
+ 8 M_{1 0 0 0 0 1 0}
+ 24 M_{0 1 0 0 0 0 1}
+ 32 M_{2 0 0 0 0 0 0}
+ 64 M_{0 0 0 0 0 0 2}
+ 78 M_{0 0 1 0 0 0 0}
\\ &+&  208 M_{0 0 0 0 0 1 0}
+ 544 M_{1 0 0 0 0 0 0}
+ 1344 M_{0 0 0 0 0 0 0}
\\
\bchi_{0 1 1 0 0 0 0}&=&
 M_{0 1 1 0 0 0 0}
+ 3 M_{1 0 0 0 1 0 0}
+ 10 M_{0 1 0 0 0 1 0}
+ 10 M_{2 0 0 0 0 0 1}
+ 30 M_{0 0 1 0 0 0 1}
+ 90 M_{1 1 0 0 0 0 0}
+ 80 M_{0 0 0 0 0 1 1}
\\ &+&   231 M_{0 0 0 0 1 0 0}
+ 570 M_{1 0 0 0 0 0 1}
+ 1344 M_{0 1 0 0 0 0 0}
+ 3024 M_{0 0 0 0 0 0 1}
\\
\bchi_{0 0 2 0 0 0 0}&=&
 M_{0 0 2 0 0 0 0}
+    M_{1 0 0 1 0 0 0}
+ 2 M_{0 1 0 0 1 0 0}
+ 3 M_{2 0 0 0 0 1 0}
+ 7 M_{0 0 1 0 0 1 0}
+ 19 M_{1 1 0 0 0 0 1}
+ 20 M_{0 0 0 0 0 2 0}
\\ &+&   46 M_{0 0 0 0 1 0 1}
+ 10 M_{3 0 0 0 0 0 0}
+ 49 M_{0 2 0 0 0 0 0}
+ 56 M_{1 0 1 0 0 0 0}
+ 104 M_{1 0 0 0 0 0 2}
+ 125 M_{0 0 0 1 0 0 0}
\\ &+&  291 M_{1 0 0 0 0 1 0}
+ 682 M_{2 0 0 0 0 0 0}
+ 638 M_{0 1 0 0 0 0 1}
+ 1338 M_{0 0 0 0 0 0 2}
+ 1402 M_{0 0 1 0 0 0 0}
+ 2908 M_{0 0 0 0 0 1 0}
\\ &+&  5938 M_{1 0 0 0 0 0 0}
+ 11844 M_{0 0 0 0 0 0 0}
\\
\bchi_{1 0 0 1 0 0 0}&=&
 M_{1 0 0 1 0 0 0}
+ 3 M_{0 1 0 0 1 0 0}
+ 4 M_{2 0 0 0 0 1 0}
+ 10 M_{0 0 1 0 0 1 0}
+ 30 M_{1 1 0 0 0 0 1}
+ 25 M_{0 0 0 0 0 2 0}
+ 75 M_{0 0 0 0 1 0 1}
\\ &+&  15 M_{3 0 0 0 0 0 0}
+ 84 M_{0 2 0 0 0 0 0}
+ 90 M_{1 0 1 0 0 0 0}
+ 180 M_{1 0 0 0 0 0 2}
+ 213 M_{0 0 0 1 0 0 0}
+ 507 M_{1 0 0 0 0 1 0}
\\ &+&  1149 M_{0 1 0 0 0 0 1}
+ 1185 M_{2 0 0 0 0 0 0}
+ 2484 M_{0 0 0 0 0 0 2}
+ 2565 M_{0 0 1 0 0 0 0}
+ 5439 M_{0 0 0 0 0 1 0}
\\ &+&  11265 M_{1 0 0 0 0 0 0}
+ 22680 M_{0 0 0 0 0 0 0}
\\
\bchi_{0 1 0 1 0 0 0}&=&
 M_{0 1 0 1 0 0 0}
+ 2 M_{0 0 1 0 1 0 0}
+ 6 M_{1 1 0 0 0 1 0}
+ 20 M_{0 2 0 0 0 0 1}
+ 15 M_{1 0 1 0 0 0 1}
+ 15 M_{0 0 0 0 1 1 0}
+ 42 M_{0 0 0 1 0 0 1}
\\ &+&   96 M_{1 0 0 0 0 1 1}
+ 40 M_{2 1 0 0 0 0 0}
+ 114 M_{0 1 1 0 0 0 0}
+ 220 M_{0 1 0 0 0 0 2}
+ 256 M_{1 0 0 0 1 0 0}
+ 565 M_{2 0 0 0 0 0 1}
\\ &+&  480 M_{0 0 0 0 0 0 3}
+ 575 M_{0 1 0 0 0 1 0}
+ 1240 M_{0 0 1 0 0 0 1}
+ 2624 M_{1 1 0 0 0 0 0}
+ 2580 M_{0 0 0 0 0 1 1}
\\ &+&   5340 M_{0 0 0 0 1 0 0}
+ 10589 M_{1 0 0 0 0 0 1}
+ 20524 M_{0 1 0 0 0 0 0}
+ 38864 M_{0 0 0 0 0 0 1}
\\
\bchi_{0 0 1 1 0 0 0}&=&
 M_{0 0 1 1 0 0 0}
+ 2 M_{1 1 0 0 1 0 0}
+ 5 M_{0 2 0 0 0 1 0}
+ 6 M_{1 0 1 0 0 1 0}
+ 5 M_{0 0 0 0 2 0 0}
+ 14 M_{0 0 0 1 0 1 0}
+ 15 M_{2 1 0 0 0 0 1}
\\ &+&  33 M_{1 0 0 0 0 2 0}
+ 37 M_{0 1 1 0 0 0 1}
+ 83 M_{1 0 0 0 1 0 1}
+ 40 M_{2 0 1 0 0 0 0}
+ 180 M_{2 0 0 0 0 0 2}
+ 94 M_{1 2 0 0 0 0 0}
\\ &+&   100 M_{0 0 2 0 0 0 0}
+ 215 M_{1 0 0 1 0 0 0}
+ 180 M_{0 1 0 0 0 1 1}
+ 467 M_{2 0 0 0 0 1 0}
+ 456 M_{0 1 0 0 1 0 0}
+ 375 M_{0 0 1 0 0 0 2}
\\ &+&   958 M_{0 0 1 0 0 1 0}
+ 750 M_{0 0 0 0 0 1 2}
+ 1964 M_{1 1 0 0 0 0 1}
+ 1920 M_{0 0 0 0 0 2 0}
+ 3963 M_{0 2 0 0 0 0 0}
+ 3850 M_{0 0 0 0 1 0 1}
\\ &+&  1010 M_{3 0 0 0 0 0 0}
+ 4005 M_{1 0 1 0 0 0 0}
+ 7374 M_{1 0 0 0 0 0 2}
+ 7700 M_{0 0 0 1 0 0 0}
+ 14642 M_{1 0 0 0 0 1 0}
\\ &+&   27546 M_{2 0 0 0 0 0 0}
+ 27263 M_{0 1 0 0 0 0 1}
+ 49698 M_{0 0 0 0 0 0 2}
+ 50206 M_{0 0 1 0 0 0 0}
+ 90408 M_{0 0 0 0 0 1 0}
\\ &+&  160642 M_{1 0 0 0 0 0 0}
+ 281268 M_{0 0 0 0 0 0 0}
\\
\bchi_{0 0 0 2 0 0 0}&=&
 M_{0 0 0 2 0 0 0}
+    M_{0 1 1 0 1 0 0}
+ 2 M_{0 0 2 0 0 1 0}
+ 2 M_{1 0 0 0 2 0 0}
+ 2 M_{1 2 0 0 0 1 0}
+ 5 M_{0 3 0 0 0 0 1}
+ 5 M_{1 0 0 1 0 1 0}
\\ &+&  11 M_{1 1 1 0 0 0 1}
+ 11 M_{0 1 0 0 1 1 0}
+ 27 M_{0 1 0 1 0 0 1}
+ 12 M_{2 0 0 0 0 2 0}
+ 25 M_{2 2 0 0 0 0 0}
+ 23 M_{0 0 1 0 0 2 0}
\\ &+&   23 M_{2 0 0 0 1 0 1}
+ 54 M_{0 0 1 0 1 0 1}
+ 25 M_{1 0 2 0 0 0 0}
+ 52 M_{2 0 0 1 0 0 0}
+ 109 M_{1 1 0 0 0 1 1}
+ 45 M_{0 0 0 0 0 3 0}
\\ &+&   64 M_{0 2 1 0 0 0 0}
+ 45 M_{3 0 0 0 0 0 2}
+ 210 M_{0 0 0 0 1 1 1}
+ 225 M_{0 2 0 0 0 0 2}
+ 210 M_{1 0 1 0 0 0 2}
+ 129 M_{0 0 1 1 0 0 0}
\\ &+&  258 M_{1 1 0 0 1 0 0}
+ 520 M_{0 2 0 0 0 1 0}
+ 408 M_{0 0 0 1 0 0 2}
+ 750 M_{1 0 0 0 0 1 2}
+ 105 M_{3 0 0 0 0 1 0}
+ 499 M_{0 0 0 0 2 0 0}
\\ &+&   501 M_{1 0 1 0 0 1 0}
+ 960 M_{2 1 0 0 0 0 1}
+ 968 M_{0 0 0 1 0 1 0}
+ 215 M_{4 0 0 0 0 0 0}
+ 1365 M_{0 1 0 0 0 0 3}
+ 1787 M_{1 0 0 0 0 2 0}
\\ &+&   1854 M_{0 1 1 0 0 0 1}
+ 3376 M_{1 0 0 0 1 0 1}
+ 1830 M_{2 0 1 0 0 0 0}
+ 3524 M_{1 2 0 0 0 0 0}
+ 6055 M_{2 0 0 0 0 0 2}
\\ &+&  3525 M_{0 0 2 0 0 0 0}
+ 6085 M_{0 1 0 0 0 1 1}
+ 10760 M_{0 0 1 0 0 0 2}
+ 6350 M_{1 0 0 1 0 0 0}
+ 11358 M_{0 1 0 0 1 0 0}
\\ &+&   11320 M_{2 0 0 0 0 1 0}
+ 2440 M_{0 0 0 0 0 0 4}
+ 18700 M_{0 0 0 0 0 1 2}
+ 20031 M_{3 0 0 0 0 0 0}
+ 19977 M_{0 0 1 0 0 1 0}
\\ &+&  34569 M_{0 0 0 0 0 2 0}
+ 34769 M_{1 1 0 0 0 0 1}
+ 60006 M_{1 0 1 0 0 0 0}
+ 60004 M_{0 2 0 0 0 0 0}
+ 59439 M_{0 0 0 0 1 0 1}
\\ &+&  101592 M_{0 0 0 1 0 0 0}
+ 100299 M_{1 0 0 0 0 0 2}
+ 170142 M_{1 0 0 0 0 1 0}
+ 281804 M_{0 1 0 0 0 0 1}
+ 461353 M_{0 0 0 0 0 0 2}
\\ &+&   282527 M_{2 0 0 0 0 0 0}
+ 462702 M_{0 0 1 0 0 0 0}
+ 750988 M_{0 0 0 0 0 1 0}
+ 1208053 M_{1 0 0 0 0 0 0}
+ 1925763 M_{0 0 0 0 0 0 0}
\\
\bchi_{1 0 0 0 1 0 0}&=&
 M_{1 0 0 0 1 0 0}
+ 4 M_{0 1 0 0 0 1 0}
+ 5 M_{2 0 0 0 0 0 1}
+ 15 M_{0 0 1 0 0 0 1}
+ 50 M_{1 1 0 0 0 0 0}
+ 44 M_{0 0 0 0 0 1 1}
+ 139 M_{0 0 0 0 1 0 0}
\\ &+& 365 M_{1 0 0 0 0 0 1}
+ 910 M_{0 1 0 0 0 0 0}
+ 2145 M_{0 0 0 0 0 0 1}
\\
\bchi_{0 1 0 0 1 0 0}&=&
 M_{0 1 0 0 1 0 0}
+ 3 M_{0 0 1 0 0 1 0}
+ 10 M_{0 0 0 0 0 2 0}
+ 10 M_{1 1 0 0 0 0 1}
+ 35 M_{0 2 0 0 0 0 0}
+ 29 M_{1 0 1 0 0 0 0}
+ 30 M_{0 0 0 0 1 0 1}
\\ &+&   88 M_{0 0 0 1 0 0 0}
+ 80 M_{1 0 0 0 0 0 2}
+ 223 M_{1 0 0 0 0 1 0}
+ 545 M_{0 1 0 0 0 0 1}
+ 1260 M_{0 0 0 0 0 0 2}
+ 538 M_{2 0 0 0 0 0 0}
\\ &+&   1262 M_{0 0 1 0 0 0 0}
+ 2800 M_{0 0 0 0 0 1 0}
+ 5976 M_{1 0 0 0 0 0 0}
+ 12341 M_{0 0 0 0 0 0 0}
\\
\bchi_{0 0 1 0 1 0 0}&=&
 M_{0 0 1 0 1 0 0}
+ 3 M_{1 1 0 0 0 1 0}
+ 9 M_{0 2 0 0 0 0 1}
+ 10 M_{1 0 1 0 0 0 1}
+ 10 M_{0 0 0 0 1 1 0}
+ 28 M_{0 0 0 1 0 0 1}
+ 72 M_{1 0 0 0 0 1 1}
\\ &+&   29 M_{2 1 0 0 0 0 0}
+ 169 M_{0 1 0 0 0 0 2}
+ 79 M_{0 1 1 0 0 0 0}
+ 196 M_{1 0 0 0 1 0 0}
+ 374 M_{0 0 0 0 0 0 3}
+ 464 M_{2 0 0 0 0 0 1}
\\ &+&  458 M_{0 1 0 0 0 1 0}
+ 1029 M_{0 0 1 0 0 0 1}
+ 2258 M_{1 1 0 0 0 0 0}
+ 2198 M_{0 0 0 0 0 1 1}
+ 4708 M_{0 0 0 0 1 0 0}
+  9574 M_{1 0 0 0 0 0 1}
\\ &+&  18998 M_{0 1 0 0 0 0 0}
+ 36774 M_{0 0 0 0 0 0 1}
\\
\bchi_{0 0 0 1 1 0 0}&=&
 M_{0 0 0 1 1 0 0}
+ 2 M_{0 1 1 0 0 1 0}
+ 6 M_{1 0 0 0 1 1 0}
+ 5 M_{0 0 2 0 0 0 1}
+ 15 M_{0 1 0 0 0 2 0}
+ 5 M_{1 2 0 0 0 0 1}
+ 14 M_{1 0 0 1 0 0 1}
\\ &+&   14 M_{0 3 0 0 0 0 0}
+ 34 M_{1 1 1 0 0 0 0}
+ 37 M_{0 1 0 0 1 0 1}
+ 91 M_{0 1 0 1 0 0 0}
+ 33 M_{2 0 0 0 0 1 1}
+ 83 M_{0 0 1 0 0 1 1}
+  180 M_{1 1 0 0 0 0 2}
\\ &+& 180 M_{0 0 0 0 0 2 1}
+ 78 M_{2 0 0 0 1 0 0}
+ 203 M_{0 0 1 0 1 0 0}
+ 437 M_{1 1 0 0 0 1 0}
+ 375 M_{0 0 0 0 1 0 2}
+ 170 M_{3 0 0 0 0 0 1}
\\ &+&  914 M_{0 0 0 0 1 1 0}
+ 750 M_{1 0 0 0 0 0 3}
+ 929 M_{0 2 0 0 0 0 1}
+ 905 M_{1 0 1 0 0 0 1}
+ 1834 M_{2 1 0 0 0 0 0}
+  1858 M_{0 0 0 1 0 0 1}
\\ &+&  3723 M_{0 1 1 0 0 0 0}
+ 3635 M_{1 0 0 0 0 1 1}
+ 7156 M_{1 0 0 0 1 0 0}
+ 6949 M_{0 1 0 0 0 0 2}
+  13480 M_{2 0 0 0 0 0 1}
+ 13524 M_{0 1 0 0 0 1 0}
\\ &+&  12954 M_{0 0 0 0 0 0 3}
+ 25015 M_{0 0 1 0 0 0 1}
+ 45599 M_{1 1 0 0 0 0 0}
+   45368 M_{0 0 0 0 0 1 1}
+ 81502 M_{0 0 0 0 1 0 0}
\\ &+&  143470 M_{1 0 0 0 0 0 1}
+ 249025 M_{0 1 0 0 0 0 0}
+ 426280 M_{0 0 0 0 0 0 1}
\\
\bchi_{0 0 0 0 2 0 0}&=&
 M_{0 0 0 0 2 0 0}
+    M_{0 0 0 1 0 1 0}
+ 2 M_{0 1 1 0 0 0 1}
+ 3 M_{1 0 0 0 0 2 0}
+ 5 M_{0 0 2 0 0 0 0}
+ 7 M_{1 0 0 0 1 0 1}
+ 19 M_{0 1 0 0 0 1 1}
\\ &+&   5 M_{1 2 0 0 0 0 0}
+ 20 M_{2 0 0 0 0 0 2}
+ 16 M_{1 0 0 1 0 0 0}
+ 46 M_{0 0 1 0 0 0 2}
+ 46 M_{0 1 0 0 1 0 0}
+ 41 M_{2 0 0 0 0 1 0}
+  110 M_{0 0 1 0 0 1 0}
\\ &+&  250 M_{1 1 0 0 0 0 1}
+ 104 M_{0 0 0 0 0 1 2}
+ 94 M_{3 0 0 0 0 0 0}
+ 254 M_{0 0 0 0 0 2 0}
+ 549 M_{0 0 0 0 1 0 1}
+   560 M_{0 2 0 0 0 0 0}
\\ &+&  1150 M_{1 0 0 0 0 0 2}
+ 539 M_{1 0 1 0 0 0 0}
+ 1159 M_{0 0 0 1 0 0 0}
+ 2362 M_{1 0 0 0 0 1 0}
+   4700 M_{0 1 0 0 0 0 1}
+ 9126 M_{0 0 0 0 0 0 2}
\\ &+&  4678 M_{2 0 0 0 0 0 0}
+ 9103 M_{0 0 1 0 0 0 0}
+ 17256 M_{0 0 0 0 0 1 0}
+   32022 M_{1 0 0 0 0 0 0}
+ 58324 M_{0 0 0 0 0 0 0}
\\
\bchi_{1 0 0 0 0 1 0}&=&
 M_{1 0 0 0 0 1 0}
+ 5 M_{0 1 0 0 0 0 1}
+ 20 M_{0 0 0 0 0 0 2}
+ 6 M_{2 0 0 0 0 0 0}
+ 21 M_{0 0 1 0 0 0 0}
+ 70 M_{0 0 0 0 0 1 0}
+   212 M_{1 0 0 0 0 0 0}
\\ &+&  588 M_{0 0 0 0 0 0 0}
\\
\bchi_{0 1 0 0 0 1 0}&=&
 M_{0 1 0 0 0 1 0}
+ 4 M_{0 0 1 0 0 0 1}
+ 16 M_{0 0 0 0 0 1 1}
+ 15 M_{1 1 0 0 0 0 0}
+ 51 M_{0 0 0 0 1 0 0}
+ 149 M_{1 0 0 0 0 0 1}
+   399 M_{0 1 0 0 0 0 0}
\\ &+&  999 M_{0 0 0 0 0 0 1}
\\
\bchi_{0 0 1 0 0 1 0}&=&
 M_{0 0 1 0 0 1 0}
+ 5 M_{0 0 0 0 0 2 0}
+ 4 M_{1 1 0 0 0 0 1}
+ 14 M_{0 2 0 0 0 0 0}
+ 15 M_{1 0 1 0 0 0 0}
+ 15 M_{0 0 0 0 1 0 1}
+ 47 M_{0 0 0 1 0 0 0}
\\ &+&   44 M_{1 0 0 0 0 0 2}
+ 133 M_{1 0 0 0 0 1 0}
+ 343 M_{0 1 0 0 0 0 1}
+ 350 M_{2 0 0 0 0 0 0}
+ 828 M_{0 0 0 0 0 0 2}
+ 845 M_{0 0 1 0 0 0 0}
\\ &+&   1957 M_{0 0 0 0 0 1 0}
+ 4366 M_{1 0 0 0 0 0 0}
+ 9387 M_{0 0 0 0 0 0 0}
\\
\bchi_{0 0 0 1 0 1 0}&=&
 M_{0 0 0 1 0 1 0}
+ 3 M_{0 1 1 0 0 0 1}
+ 4 M_{1 0 0 0 0 2 0}
+ 9 M_{0 0 2 0 0 0 0}
+ 10 M_{1 0 0 0 1 0 1}
+ 30 M_{0 1 0 0 0 1 1}
+ 9 M_{1 2 0 0 0 0 0}
\\ &+&   25 M_{2 0 0 0 0 0 2}
+ 27 M_{1 0 0 1 0 0 0}
+ 75 M_{0 0 1 0 0 0 2}
+ 78 M_{0 1 0 0 1 0 0}
+ 69 M_{2 0 0 0 0 1 0}
+ 193 M_{0 0 1 0 0 1 0}
+  449 M_{1 1 0 0 0 0 1}
\\ &+&  180 M_{0 0 0 0 0 1 2}
+ 165 M_{3 0 0 0 0 0 0}
+ 460 M_{0 0 0 0 0 2 0}
+ 1014 M_{0 0 0 0 1 0 1}
+ 1008 M_{0 2 0 0 0 0 0}
+   2169 M_{1 0 0 0 0 0 2}
\\ &+&  999 M_{1 0 1 0 0 0 0}
+ 2184 M_{0 0 0 1 0 0 0}
+ 4549 M_{1 0 0 0 0 1 0}
+ 9198 M_{0 1 0 0 0 0 1}
+  18063 M_{0 0 0 0 0 0 2}
+ 9189 M_{2 0 0 0 0 0 0}
\\ &+&  18114 M_{0 0 1 0 0 0 0}
+ 34807 M_{0 0 0 0 0 1 0}
+ 65475 M_{1 0 0 0 0 0 0}
+   120771 M_{0 0 0 0 0 0 0}
\\
\bchi_{0 0 0 0 1 1 0}&=&
 M_{0 0 0 0 1 1 0}
+ 2 M_{0 0 0 1 0 0 1}
+ 8 M_{1 0 0 0 0 1 1}
+ 5 M_{0 1 1 0 0 0 0}
+ 24 M_{0 1 0 0 0 0 2}
+ 19 M_{1 0 0 0 1 0 0}
+ 59 M_{0 1 0 0 0 1 0}
\\ &+&   64 M_{0 0 0 0 0 0 3}
+ 54 M_{2 0 0 0 0 0 1}
+ 154 M_{0 0 1 0 0 0 1}
+ 374 M_{1 1 0 0 0 0 0}
+ 384 M_{0 0 0 0 0 1 1}
+ 879 M_{0 0 0 0 1 0 0}
\\ &+&   1958 M_{1 0 0 0 0 0 1}
+ 4193 M_{0 1 0 0 0 0 0}
+ 8694 M_{0 0 0 0 0 0 1}
\\
\bchi_{0 0 0 0 0 2 0}&=&
 M_{0 0 0 0 0 2 0}
+    M_{0 0 0 0 1 0 1}
+ 2 M_{0 0 0 1 0 0 0}
+ 4 M_{1 0 0 0 0 0 2}
+ 9 M_{1 0 0 0 0 1 0}
+ 29 M_{0 1 0 0 0 0 1}
+ 30 M_{2 0 0 0 0 0 0}
\\ &+&   84 M_{0 0 0 0 0 0 2}
+ 80 M_{0 0 1 0 0 0 0}
+ 209 M_{0 0 0 0 0 1 0}
+ 510 M_{1 0 0 0 0 0 0}
+ 1197 M_{0 0 0 0 0 0 0}
\\
\bchi_{1 0 0 0 0 0 1}&=&
 M_{1 0 0 0 0 0 1}
+ 6 M_{0 1 0 0 0 0 0}
+ 27 M_{0 0 0 0 0 0 1}
\\
\bchi_{0 1 0 0 0 0 1}&=&
 M_{0 1 0 0 0 0 1}
+ 5 M_{0 0 1 0 0 0 0}
+ 6 M_{0 0 0 0 0 0 2}
+ 22 M_{0 0 0 0 0 1 0}
+ 75 M_{1 0 0 0 0 0 0}
+ 225 M_{0 0 0 0 0 0 0}
\\
\bchi_{0 0 1 0 0 0 1}&=&
 M_{0 0 1 0 0 0 1}
+ 5 M_{0 0 0 0 0 1 1}
+ 5 M_{1 1 0 0 0 0 0}
+ 20 M_{0 0 0 0 1 0 0}
+ 66 M_{1 0 0 0 0 0 1}
+ 196 M_{0 1 0 0 0 0 0}
+   531 M_{0 0 0 0 0 0 1}
\\
\bchi_{0 0 0 1 0 0 1}&=&
 M_{0 0 0 1 0 0 1}
+ 4 M_{1 0 0 0 0 1 1}
+ 4 M_{0 1 1 0 0 0 0}
+ 15 M_{0 1 0 0 0 0 2}
+ 14 M_{1 0 0 0 1 0 0}
+ 45 M_{0 1 0 0 0 1 0}
\\ &+&   45 M_{0 0 0 0 0 0 3}
+ 40 M_{2 0 0 0 0 0 1}
+ 125 M_{0 0 1 0 0 0 1}
+ 319 M_{1 1 0 0 0 0 0}
+ 325 M_{0 0 0 0 0 1 1}
+ 784 M_{0 0 0 0 1 0 0}
\\ &+&   1809 M_{1 0 0 0 0 0 1}
+ 4004 M_{0 1 0 0 0 0 0}
+ 8529 M_{0 0 0 0 0 0 1}
\\
\bchi_{0 0 0 0 1 0 1}&=&
 M_{0 0 0 0 1 0 1}
+ 3 M_{0 0 0 1 0 0 0}
+ 5 M_{1 0 0 0 0 0 2}
+ 13 M_{1 0 0 0 0 1 0}
+ 45 M_{0 1 0 0 0 0 1}
+ 39 M_{2 0 0 0 0 0 0}
+   135 M_{0 0 0 0 0 0 2}
\\ &+&  129 M_{0 0 1 0 0 0 0}
+ 351 M_{0 0 0 0 0 1 0}
+ 879 M_{1 0 0 0 0 0 0}
+ 2079 M_{0 0 0 0 0 0 0}
\\
\bchi_{0 0 0 0 0 1 1}&=&
 M_{0 0 0 0 0 1 1}
+ 2 M_{0 0 0 0 1 0 0}
+ 10 M_{1 0 0 0 0 0 1}
+ 35 M_{0 1 0 0 0 0 0}
+ 111 M_{0 0 0 0 0 0 1}
\\
\bchi_{0 0 0 0 0 0 2}&=&
 M_{0 0 0 0 0 0 2}
+    M_{0 0 0 0 0 1 0}
+ 5 M_{1 0 0 0 0 0 0}
+ 21 M_{0 0 0 0 0 0 0}
\end{eqnarray*}
\subsection{First order polynomials}
The iterative methods given in \cite{fgp05c} allow us to solve the Schr\"{o}dinger equation (\ref{sch}) for general $\kappa$. The eigenfuntions are polynomials. In this and the next subsection, we present a partial list of such polynomials of first and second order.
\begin{eqnarray*}
P^\kappa_{1000000}(z)&=&z_1
+\frac{ 7 (-1 + \kappa) }{ 1 + 17 \kappa }
\\
P^\kappa_{0100000}(z)&=&z_2
+\frac{ 6 (-1 + \kappa) z_7 }{ 1 + 11 \kappa }
\\
P^\kappa_{0010000}(z)&=&z_3
+\frac{ 5 (-1 + \kappa) z_6 }{ 1 + 7 \kappa }
+\frac{ 8 (-1 + \kappa) (-1 + 8 \kappa) z_1 }{ (1 + 7 \kappa) (1 + 8 \kappa) }
+\frac{ 2 (-1 + \kappa) (-1 - 159 \kappa + 136 \kappa^2) }{ (1 + 7 \kappa) (1 + 8 \kappa) 
   (1 + 11 \kappa) }
\\
P^\kappa_{0001000}(z)&=&z_4
+\frac{ 4 (-1 + \kappa) z_1 z_6 }{ 1 + 5 \kappa }
+\frac{ 3 (-1 + \kappa) (-3 + 5 \kappa) z_2 z_7 }{ (1 + 5 \kappa)^2 }
+\frac{ -9 (-1 + \kappa) (1 + 22 \kappa + 5 \kappa^2) z_7^2 }{ (1 + 5 \kappa)^2 (1 + 7 \kappa) }
\\ &+&  \frac{ 9 (-1 + \kappa) (-1 + 3 \kappa) z_1^2 }{ (1 + 5 \kappa) (1 + 7 \kappa) }
+\frac{ -2 (-1 + \kappa) (-7 + 11 \kappa + 385 \kappa^2 + 175 \kappa^3) z_3 }{ 
  (1 + 5 \kappa)^3 (1 + 7 \kappa) }
\\ &+&  \frac{ (-1 + \kappa) (78 + 1255 \kappa + 3653 \kappa^2 - 6295 \kappa^3 + 3325 \kappa^4) z_6 }{ 
  (1 + 5 \kappa)^3 (1 + 7 \kappa) (2 + 11 \kappa) }
\\ &+&  \frac{ -2 (-1 + \kappa) (-22 - 755 \kappa - 3477 \kappa^2 + 11255 \kappa^3 + 175 \kappa^4) z_1 }{ 
  (1 + 5 \kappa)^3 (1 + 7 \kappa) (2 + 11 \kappa) }
\\ &+&  \frac{ 2 (-1 + \kappa) (18 + 365 \kappa + 8123 \kappa^2 - 2045 \kappa^3 + 2275 \kappa^4) }{ 
  (1 + 5 \kappa)^3 (1 + 7 \kappa) (2 + 11 \kappa) }
\\
P^\kappa_{0000100}(z)&=&z_5
+\frac{ 5 (-1 + \kappa) z_1 z_7 }{ 1 + 7 \kappa }
+\frac{ 7 (-1 + \kappa) (-4 + 7 \kappa) z_2 }{ (1 + 7 \kappa) (2 + 13 \kappa) }
+\frac{ 5 (-1 + \kappa) (-6 - 137 \kappa + 56 \kappa^2) z_7 }{ 
  (1 + 7 \kappa) (1 + 8 \kappa) (2 + 13 \kappa) }
\\
P^\kappa_{0000010}(z)&=&z_6
+\frac{ 6 (-1 + \kappa) z_1 }{ 1 + 9 \kappa }
+\frac{ 15 (-1 + \kappa) (-1 + 5 \kappa) }{ (1 + 9 \kappa) (1 + 13 \kappa) }
\\
P^\kappa_{0000001}(z)&=&z_7
\end{eqnarray*}
\subsection{Second order polynomials}
\begin{eqnarray*}
P^\kappa_{2000000}(z)&=&z_1^2
+\frac{ -2 z_3 }{ 1 + \kappa }
+\frac{ -10 \kappa z_6 }{ (1 + \kappa) (1 + 4 \kappa) }
+\frac{ 2 (-3 - 6 \kappa - 119 \kappa^2 + 28 \kappa^3) z_1 }{ (1 + \kappa) (1 + 4 \kappa) 
   (3 + 17 \kappa) }
\\ &+&  \frac{ -42 - 459 \kappa - 290 \kappa^2 - 3205 \kappa^3 + 196 \kappa^4 }{ 
  (1 + \kappa) (1 + 4 \kappa) (2 + 17 \kappa) (3 + 17 \kappa) }
\\
P^\kappa_{1100000}(z)&=&z_1 z_2
+\frac{ -5 z_5 }{ 1 + 4 \kappa }
+\frac{ (6 - 95 \kappa + 24 \kappa^2) z_1 z_7 }{ (1 + 4 \kappa) (2 + 11 \kappa) }
+\frac{ 28 (-1 + \kappa) \kappa (-26 + 11 \kappa) z_2 }{ (1 + 4 \kappa) (2 + 11 \kappa) 
   (3 + 17 \kappa) }
\\ &+&  \frac{ (-1 + \kappa) (138 + 365 \kappa - 9979 \kappa^2 + 1176 \kappa^3) z_7 }{ 
  (1 + 4 \kappa) (1 + 7 \kappa) (2 + 11 \kappa) (3 + 17 \kappa) }
\\
P^\kappa_{0200000}(z)&=&z_2^2
+\frac{ -2 z_4 }{ 1 + \kappa }
+\frac{ -8 \kappa z_1 z_6 }{ (1 + \kappa) (1 + 3 \kappa) }
+\frac{ 6 (-1 + \kappa) (1 - \kappa + 6 \kappa^2) z_2 z_7 }{ (1 + \kappa) (1 + 3 \kappa) 
   (3 + 11 \kappa) }
+  \frac{ -2 (1 + 23 \kappa^2) z_1^2 }{ (1 + \kappa) (1 + 3 \kappa) (1 + 5 \kappa) }
\\ &+& \frac{ 18 (-1 + \kappa) (2 + 13 \kappa - 7 \kappa^2 + 6 \kappa^3) z_7^2 }{ 
  (1 + \kappa) (1 + 3 \kappa) (2 + 11 \kappa) (3 + 11 \kappa) }
+  \frac{ 4 (3 + 23 \kappa + 141 \kappa^2 + 493 \kappa^3 + 180 \kappa^4) z_3 }{ 
  (1 + \kappa) (1 + 3 \kappa) (1 + 4 \kappa) (1 + 5 \kappa) (3 + 11 \kappa) }
\\ &+&  \frac{ -2 (-42 - 537 \kappa - 2397 \kappa^2 - 6715 \kappa^3 - 14529 \kappa^4 + 2380 \kappa^5) z_6 }{ 
  (1 + \kappa) (1 + 3 \kappa) (1 + 4 \kappa) (1 + 5 \kappa) (2 + 11 \kappa) (3 + 11 \kappa) }
\\ &+&  \frac{ -4 (-6 - 99 \kappa - 205 \kappa^2 - 1021 \kappa^3 - 12161 \kappa^4 + 2572 \kappa^5) z_1 }{ 
  (1 + \kappa) (1 + 3 \kappa) (1 + 4 \kappa) (1 + 5 \kappa) (2 + 11 \kappa) (3 + 11 \kappa) }
\\ &+&  \frac{ -4 (-1 + \kappa) (18 + 325 \kappa + 2143 \kappa^2 + 2067 \kappa^3 - 14045 \kappa^4 + 
     22012 \kappa^5) }{ (1 + \kappa) (1 + 3 \kappa) (1 + 4 \kappa) (1 + 5 \kappa) (1 + 7 \kappa) 
   (2 + 11 \kappa) (3 + 11 \kappa) }
\\
P^\kappa_{1010000}(z)&=&z_1 z_3
+\frac{ -3 z_4 }{ 1 + 2 \kappa }
+\frac{ (-2 - 35 \kappa + 10 \kappa^2) z_1 z_6 }{ (1 + 2 \kappa) (2 + 7 \kappa) }
+\frac{ -6 (-1 + \kappa) (2 + 15 \kappa) z_2 z_7 }{ (1 + 2 \kappa) (1 + 4 \kappa) (2 + 7 \kappa) }
\\ &+&  \frac{ (-18 - 47 \kappa - 704 \kappa^2 + 256 \kappa^3) z_1^2 }{ 
  (1 + 2 \kappa) (2 + 7 \kappa) (3 + 16 \kappa) }
+\frac{ 9 (-2 + 17 \kappa) z_7^2 }{ (1 + 2 \kappa) (1 + 4 \kappa) (2 + 7 \kappa) }
\\ &+&  \frac{ (-216 - 2054 \kappa + 1429 \kappa^2 + 33210 \kappa^3 + 15224 \kappa^4 + 6272 \kappa^5) z_3 }{ 
  (1 + 2 \kappa) (1 + 4 \kappa) (2 + 7 \kappa) (3 + 16 \kappa) (4 + 17 \kappa) }
\\ &+&  \frac{ 2 (-1 + \kappa) (-48 + 1190 \kappa + 6697 \kappa^2 - 9260 \kappa^3 + 2240 \kappa^4) z_6 }{ 
  (1 + 2 \kappa) (1 + 4 \kappa) (2 + 7 \kappa) (3 + 16 \kappa) (4 + 17 \kappa) }
\\ &+&  \frac{ 12 (-1 + \kappa) (-80 - 100 \kappa + 7708 \kappa^2 + 27189 \kappa^3 - 30716 \kappa^4 + 
     12736 \kappa^5) z_1 }{ (1 + 2 \kappa) (1 + 4 \kappa) (2 + 7 \kappa) (2 + 11 \kappa) (3 + 16 \kappa) 
   (4 + 17 \kappa) }
\\ &+&  \frac{ 4 (-1 + \kappa) (-384 - 6676 \kappa - 12672 \kappa^2 + 67253 \kappa^3 - 75612 \kappa^4 + 
     7616 \kappa^5) }{ (1 + 2 \kappa) (1 + 4 \kappa) (2 + 7 \kappa) (2 + 11 \kappa) (3 + 16 \kappa) 
   (4 + 17 \kappa) }
\\
P^\kappa_{0110000}(z)&=&z_2 z_3
+\frac{ -4 z_1 z_5 }{ 1 + 3 \kappa }
+\frac{ 5 (-1 + \kappa) (-2 + 3 \kappa) z_2 z_6 }{ (1 + 3 \kappa) (2 + 7 \kappa) }
+\frac{ -4 (-1 + 7 \kappa) z_1^2 z_7 }{ (1 + 3 \kappa) (1 + 5 \kappa) }
\\ &+&  \frac{ 6 (-4 + 51 \kappa + 311 \kappa^2 + 41 \kappa^3 + 105 \kappa^4) z_3 z_7 }{ 
  (1 + 3 \kappa) (1 + 5 \kappa) (2 + 7 \kappa) (3 + 11 \kappa) }
\\ &+&  \frac{ 2 (-1 + \kappa) (51 + 397 \kappa - 6 \kappa^2 - 2992 \kappa^3 + 2640 \kappa^4) z_1 z_2 }{ 
  (1 + 3 \kappa) (1 + 4 \kappa) (1 + 5 \kappa) (2 + 7 \kappa) (3 + 11 \kappa) }
\\ &+&  \frac{ 6 (-16 + 17 \kappa + 673 \kappa^2 - 245 \kappa^3 + 75 \kappa^4) z_6 z_7 }{ 
  (1 + 3 \kappa) (1 + 5 \kappa) (2 + 7 \kappa) (3 + 11 \kappa) }
\\ &+&  \frac{ 2 (-1 + \kappa) (-123 - 821 \kappa + 1018 \kappa^2 + 10196 \kappa^3 + 1280 \kappa^4) z_5 }{ 
  (1 + 3 \kappa) (1 + 4 \kappa)^2 (1 + 5 \kappa) (2 + 7 \kappa) (3 + 11 \kappa) }
\\ &+&  \frac{ 6 (13 - 156 \kappa - 2853 \kappa^2 - 2376 \kappa^3 + 27380 \kappa^4 - 11328 \kappa^5 + 
     1920 \kappa^6) z_1 z_7 }{ (1 + 3 \kappa) (1 + 4 \kappa)^2 (1 + 5 \kappa) (2 + 7 \kappa) 
   (3 + 11 \kappa) }
\\ &+&  \frac{ 6 (-1 + \kappa) (-12 - 595 \kappa - 2610 \kappa^2 + 7325 \kappa^3 - 2248 \kappa^4 + 1360 \kappa^5) 
   z_2 }{ (1 + 3 \kappa) (1 + 4 \kappa)^2 (1 + 5 \kappa) (2 + 7 \kappa) (3 + 11 \kappa) }
\\ &+&  \frac{ 2 (-1 + \kappa) (-42 - 3713 \kappa - 46855 \kappa^2 - 49890 \kappa^3 + 620062 \kappa^4 - 
     178192 \kappa^5 + 24480 \kappa^6) z_7 }{ 
  (1 + 3 \kappa) (1 + 4 \kappa)^2 (1 + 5 \kappa) (2 + 7 \kappa) (2 + 11 \kappa) (3 + 11 \kappa) }
\\
P^\kappa_{1000100}(z)&=&z_1 z_5
+\frac{ -5 z_2 z_6 }{ 1 + 4 \kappa }
+\frac{ 5 (-1 + \kappa) z_1^2 z_7 }{ 1 + 7 \kappa }
+\frac{ -15 (-1 + \kappa) (1 + 6 \kappa) z_3 z_7 }{ (1 + 4 \kappa)^2 (1 + 7 \kappa) }
\\ &+&  \frac{ (-1 + \kappa) (-27 - 320 \kappa + 112 \kappa^2) z_1 z_2 }{ (1 + 4 \kappa)^2 
   (3 + 13 \kappa) }
+  \frac{ -(-1 + \kappa) (19 + 309 \kappa + 1182 \kappa^2) z_6 z_7}{ 
  (1 + 4 \kappa)^2 (1 + 5 \kappa) (1 + 7 \kappa) }
\\ &+&  \frac{ (-1 + \kappa) (648 + 8119 \kappa + 26227 \kappa^2 + 12296 \kappa^3 + 7280 \kappa^4) z_5 }{ 
  (1 + 4 \kappa)^2 (1 + 5 \kappa) (3 + 13 \kappa) (4 + 17 \kappa) }
\\ &+&  \frac{ 2 (-174 - 4381 \kappa - 20915 \kappa^2 + 61085 \kappa^3 + 363609 \kappa^4 - 239624 \kappa^5 + 
     42000 \kappa^6) z_1 z_7 }{ (1 + 4 \kappa)^2 (1 + 5 \kappa) (1 + 7 \kappa) (3 + 13 \kappa) 
   (4 + 17 \kappa) }
\\ &+&  \frac{ 10 (36 + 449 \kappa - 4256 \kappa^2 - 63640 \kappa^3 - 162486 \kappa^4 + 81791 \kappa^5 - 
     107534 \kappa^6 + 13720 \kappa^7) z_2 }{ 
  (1 + 4 \kappa)^2 (1 + 5 \kappa)^2 (1 + 7 \kappa) (3 + 13 \kappa) (4 + 17 \kappa) }
\\ &+&  \frac{ (-1 + \kappa) (-672 - 17905 \kappa - 81245 \kappa^2 + 497285 \kappa^3 + 2671037 \kappa^4 - 
     1392420 \kappa^5 + 98000 \kappa^6) z_7 }{ 
  (1 + 4 \kappa)^2 (1 + 5 \kappa)^2 (1 + 7 \kappa) (3 + 13 \kappa) (4 + 17 \kappa) }
\\
P^\kappa_{1000010}(z)&=&z_1 z_6
+\frac{ -6 z_2 z_7 }{ 1 + 5 \kappa }
+\frac{ -9 (-1 + 7 \kappa) z_7^2 }{ (1 + 5 \kappa) (1 + 8 \kappa) }
+\frac{ 6 (-1 + \kappa) z_1^2 }{ 1 + 9 \kappa }
+\frac{ -3 (-1 + \kappa) (7 + 55 \kappa) z_3 }{ (1 + 5 \kappa)^2 (1 + 9 \kappa) }
\\ &+&  \frac{ (18 + 1213 \kappa + 15375 \kappa^2 + 51579 \kappa^3 - 15985 \kappa^4 + 12600 \kappa^5) z_6 }{ 
  (1 + 5 \kappa)^2 (1 + 8 \kappa) (1 + 9 \kappa) (3 + 17 \kappa) }
\\&+&\frac{ 3 (-54 - 775 \kappa + 4734 \kappa^2 + 107248 \kappa^3 + 344272 \kappa^4 - 252825 \kappa^5 + 
     121400 \kappa^6) z_1 }{ (1 + 5 \kappa)^2 (1 + 8 \kappa) (1 + 9 \kappa) (2 + 13 \kappa) 
   (3 + 17 \kappa) }
\\ &+&  \frac{ 3 (54 - 245 \kappa - 9444 \kappa^2 + 22702 \kappa^3 + 391238 \kappa^4 - 115305 \kappa^5 + 
     35000 \kappa^6) }{ (1 + 5 \kappa)^2 (1 + 8 \kappa) (1 + 9 \kappa) (2 + 13 \kappa) 
   (3 + 17 \kappa) }
\\
P^\kappa_{0100010}(z)&=&z_2 z_6
+\frac{ -5 z_3 z_7 }{ 1 + 4 \kappa }
+\frac{ (6 - 95 \kappa + 24 \kappa^2) z_6 z_7 }{ (1 + 4 \kappa) (2 + 11 \kappa) }
+\frac{ 6 (-1 + \kappa) (-3 + 4 \kappa) z_1 z_2 }{ (1 + 4 \kappa) (2 + 9 \kappa) }
\\ &+&  \frac{ 2 (-42 + 643 \kappa + 4519 \kappa^2 + 600 \kappa^3) z_5 }{ 
  (1 + 4 \kappa) (2 + 9 \kappa) (2 + 11 \kappa) (3 + 13 \kappa) }
\\ &+&  \frac{ 2 (-1 + \kappa) (114 + 113 \kappa - 11559 \kappa^2 - 46218 \kappa^3 + 4680 \kappa^4) z_1 z_7 }{  (1 + 4 \kappa) (1 + 5 \kappa) (2 + 9 \kappa) (2 + 11 \kappa) (3 + 13 \kappa) }
\\ &+&  \frac{ (180 - 3864 \kappa - 20509 \kappa^2 + 80848 \kappa^3 - 4515 \kappa^4 + 16500 \kappa^5) z_2 }{ 
  (1 + 4 \kappa) (1 + 5 \kappa) (2 + 9 \kappa) (2 + 11 \kappa) (3 + 13 \kappa) }
\\ &+&  \frac{ 2 (-1 + \kappa) (-42 + 2347 \kappa + 38355 \kappa^2 - 27714 \kappa^3 + 4500 \kappa^4) z_7 }{ 
  (1 + 4 \kappa) (1 + 5 \kappa) (2 + 9 \kappa) (2 + 11 \kappa) (3 + 13 \kappa) }
\\
P^\kappa_{0000020}(z)&=&z_6^2
+\frac{ -2 z_5 z_7 }{ 1 + \kappa }
+\frac{ -2 (-1 + \kappa) z_4 }{ (1 + \kappa) (1 + 2 \kappa) }
+\frac{ -10 \kappa z_1 z_7^2 }{ (1 + \kappa) (1 + 4 \kappa) }
+\frac{ 4 \kappa (13 + 43 \kappa + 10 \kappa^2 + 24 \kappa^3) z_1 z_6 }{ 
  3 (1 + \kappa) (1 + 2 \kappa) (1 + 3 \kappa) (1 + 4 \kappa) }
\\ &+&  \frac{ -2 (-1 + \kappa) \kappa (7 + 36 \kappa) z_2 z_7 }{ 
  (1 + \kappa) (1 + 2 \kappa) (1 + 3 \kappa) (1 + 4 \kappa) }
+ \frac{ 2 (-1 + \kappa) (6 + 47 \kappa + 59 \kappa^2 - 88 \kappa^3 + 48 \kappa^4) z_1^2 }{ 
  (1 + \kappa) (1 + 2 \kappa) (1 + 3 \kappa) (1 + 4 \kappa) (2 + 9 \kappa) }
\\ &+&  \frac{ -2 (-1 + \kappa) (-1 - 8 \kappa + 30 \kappa^2) z_7^2 }{ 
  (1 + \kappa) (1 + 2 \kappa) (1 + 3 \kappa) (1 + 4 \kappa) }
+\frac{ 4 (6 + 29 \kappa + 36 \kappa^2 + 199 \kappa^3 + 60 \kappa^4) z_3 }{ 
  (1 + \kappa) (1 + 2 \kappa) (1 + 3 \kappa) (1 + 4 \kappa) (2 + 9 \kappa) }
\\ &+&  \frac{ 4 (18 + 213 \kappa - 140 \kappa^2 - 1100 \kappa^3 + 11362 \kappa^4 + 2787 \kappa^5 + 2700 \kappa^6) 
   z_6 }{ 3 (1 + \kappa) (1 + 2 \kappa) (1 + 3 \kappa) (1 + 4 \kappa) (2 + 9 \kappa) (3 + 13 \kappa) }
\\ &+&  \frac{ 4 (18 + 285 \kappa + 949 \kappa^2 - 2675 \kappa^3 - 5493 \kappa^4 + 33950 \kappa^5 + 
     1646 \kappa^6 + 3000 \kappa^7) z_1 }{ 
  (1 + \kappa) (1 + 2 \kappa) (1 + 3 \kappa) (1 + 4 \kappa) (1 + 5 \kappa) (2 + 9 \kappa) (3 + 13 \kappa) }
\\ &+&  \frac{ (-1 + \kappa) (-204 - 4208 \kappa - 37487 \kappa^2 - 140165 \kappa^3 - 24391 \kappa^4 + 
     655613 \kappa^5 - 66238 \kappa^6 + 75000 \kappa^7) }{ 
  (1 + \kappa) (1 + 2 \kappa) (1 + 3 \kappa) (1 + 4 \kappa) (1 + 5 \kappa) (2 + 9 \kappa) (2 + 13 \kappa) 
   (3 + 13 \kappa) }
\\
P^\kappa_{1000001}(z)&=&z_1 z_7
+\frac{ -7 z_2 }{ 1 + 6 \kappa }
+\frac{ (16 - 191 \kappa + 42 \kappa^2) z_7 }{ (1 + 6 \kappa) (2 + 17 \kappa) }
\\
P^\kappa_{0100001}(z)&=&z_2 z_7
+\frac{ -6 z_3 }{ 1 + 5 \kappa }
+\frac{ 6 (-1 + \kappa) z_7^2 }{ 1 + 11 \kappa }
+\frac{ -2 (-7 - 31 \kappa + 290 \kappa^2) z_6 }{ (1 + 5 \kappa) (1 + 6 \kappa) (1 + 11 \kappa) }
\\ &+&  \frac{ -16 (-1 - 26 \kappa - 119 \kappa^2 + 398 \kappa^3) z_1 }{ 
  (1 + 5 \kappa) (1 + 6 \kappa) (1 + 7 \kappa) (1 + 11 \kappa) }
+\frac{ -4 (-1 + \kappa) (-7 - 51 \kappa + 826 \kappa^2) }{ 
  (1 + 5 \kappa) (1 + 6 \kappa) (1 + 7 \kappa) (1 + 11 \kappa) }
\\
P^\kappa_{0010001}(z)&=&z_3 z_7
+\frac{ 5 (-1 + \kappa) z_6 z_7 }{ 1 + 7 \kappa }
+\frac{ -6 z_1 z_2 }{ 1 + 5 \kappa }
+\frac{ -10 (-1 + \kappa) (4 + 25 \kappa) z_5 }{ (1 + 5 \kappa) (1 + 7 \kappa) (2 + 9 \kappa) }
\\ &+&  \frac{ 2 (-1 + \kappa) (-72 - 1183 \kappa - 4188 \kappa^2 + 2880 \kappa^3) z_1 z_7 }{ 
  (1 + 5 \kappa) (1 + 7 \kappa) (2 + 9 \kappa) (3 + 16 \kappa) }
\\ &+&  \frac{ -28 (-1 + \kappa) (-21 - 260 \kappa - 333 \kappa^2 + 3056 \kappa^3) z_2 }{ 
  (1 + 5 \kappa) (1 + 7 \kappa) (2 + 9 \kappa) (2 + 11 \kappa) (3 + 16 \kappa) }
\\ &+&  \frac{ 2 (-72 - 4430 \kappa - 9270 \kappa^2 + 121301 \kappa^3 - 54561 \kappa^4 + 12240 \kappa^5) 
   z_7 }{ (1 + 5 \kappa) (1 + 7 \kappa) (2 + 9 \kappa) (2 + 11 \kappa) (3 + 16 \kappa) }
\\
P^\kappa_{0000101}(z)&=&z_5 z_7
+\frac{ -4 z_4 }{ 1 + 3 \kappa }
+\frac{ 5 (-1 + \kappa) z_1 z_7^2 }{ 1 + 7 \kappa }
+\frac{ -8 (-1 - \kappa + 22 \kappa^2) z_1 z_6 }{ (1 + 3 \kappa) (1 + 4 \kappa) (1 + 7 \kappa) }
\\ &+&  \frac{ (-1 + \kappa) (-33 - 239 \kappa + 84 \kappa^2) z_2 z_7 }{ 
  (1 + 3 \kappa) (1 + 4 \kappa) (3 + 13 \kappa) }
+\frac{ -12 (-1 + \kappa) (1 + 12 \kappa) z_1^2 }{ (1 + 3 \kappa) (1 + 4 \kappa) (1 + 7 \kappa) }
\\ &+&  \frac{ (-9 + 327 \kappa + 3089 \kappa^2 - 1267 \kappa^3 + 420 \kappa^4) z_7^2 }{ 
  (1 + 3 \kappa) (1 + 4 \kappa) (1 + 7 \kappa) (3 + 13 \kappa) }
+\frac{ -12 (-1 + \kappa) (-21 - 118 \kappa + 8 \kappa^2) z_3 }{ 
  (1 + 3 \kappa) (1 + 4 \kappa) (2 + 9 \kappa) (3 + 13 \kappa) }
\\ &+&  \frac{ -4 (27 + 1050 \kappa + 8824 \kappa^2 + 18278 \kappa^3 - 11811 \kappa^4 + 25872 \kappa^5) z_6  }{ (1 + 3 \kappa) (1 + 4 \kappa) (1 + 5 \kappa) (1 + 7 \kappa) (2 + 9 \kappa) (3 + 13 \kappa) }
\\ &+&  \frac{ -48 (-1 + \kappa) \kappa (-89 - 1205 \kappa - 3637 \kappa^2 + 2159 \kappa^3) z_1 }{ 
  (1 + 3 \kappa) (1 + 4 \kappa) (1 + 5 \kappa) (1 + 7 \kappa) (2 + 9 \kappa) (3 + 13 \kappa) }
\\ &+&  \frac{ -8 (3 - 203 \kappa + 3439 \kappa^2 + 24833 \kappa^3 - 17242 \kappa^4 + 10290 \kappa^5) }{ 
  (1 + 3 \kappa) (1 + 4 \kappa) (1 + 5 \kappa) (1 + 7 \kappa) (2 + 9 \kappa) (3 + 13 \kappa) }
\\
P^\kappa_{0000011}(z)&=&z_6 z_7
+\frac{ -3 z_5 }{ 1 + 2 \kappa }
+\frac{ (-2 - 43 \kappa + 12 \kappa^2) z_1 z_7 }{ (1 + 2 \kappa) (2 + 9 \kappa) }
+\frac{ -7 (-1 + \kappa) (2 + 19 \kappa) z_2 }{ (1 + 2 \kappa) (1 + 5 \kappa) (2 + 9 \kappa) }
\\ &+&  \frac{ 2 (-1 + \kappa) (22 - 2 \kappa - 1052 \kappa^2 + 375 \kappa^3) z_7 }{ 
  (1 + 2 \kappa) (1 + 5 \kappa) (2 + 9 \kappa) (2 + 13 \kappa) }
\\
P^\kappa_{0000002}(z)&=&z_7^2
+\frac{ -2 z_6 }{ 1 + \kappa }
+\frac{ -12 \kappa z_1 }{ (1 + \kappa) (1 + 5 \kappa) }
+\frac{ -2 (1 + 59 \kappa^2) }{ (1 + \kappa) (1 + 5 \kappa) (1 + 9 \kappa) }
\end{eqnarray*}
\subsection{Generalized quadratic Clebsch-Gordan series}
For each $\kappa$, the product of polynomials can be decomposed as a linear combination of polynomials of the same $\kappa$. The terms entering in this decomposition are exactly the same entering in the product of characters, i.e. in the corresponding Clebsch-Gordan series, while the coefficients are rational functions of $\kappa$. The method for computing these coefficients was explained in \cite{fgp05c}. Here we give some of  the quadratic Clebsch-Gordan series for general $\kappa$.
\begin{eqnarray*}
P^\kappa_{1000000}\times P^\kappa_{1000000}&=&
 P^\kappa_{2000000}
+\frac{ 2 }{ 1 + \kappa } P^\kappa_{0010000}
+\frac{ 10 (1 + 3 \kappa) }{ (1 + 4 \kappa) (1 + 7 \kappa) } P^\kappa_{0000010}
\\ &+&  \frac{ 24 (4 + 103 \kappa + 547 \kappa^2 + 696 \kappa^3) }{ 
  (1 + 8 \kappa) (1 + 9 \kappa) (1 + 17 \kappa) (3 + 17 \kappa) } P^\kappa_{1000000}
\\ &+&  \frac{ 252 (1 + 3 \kappa) (1 + 5 \kappa) (1 + 8 \kappa) (1 + 18 \kappa) }{ 
  (1 + 11 \kappa) (1 + 13 \kappa) (1 + 17 \kappa)^2 (2 + 17 \kappa) } 
\\
P^\kappa_{1000000}\times P^\kappa_{0100000}&=&
 P^\kappa_{1100000}
+\frac{ 5 }{ 1 + 4 \kappa } P^\kappa_{0000100}
+\frac{ 32 (1 + 2 \kappa) (1 + 12 \kappa) }{ (1 + 7 \kappa) (1 + 11 \kappa) (2 + 11 \kappa) } P^\kappa_{1000001}
\\ &+&  \frac{ 42 (1 + 3 \kappa) (1 + 14 \kappa) (5 + 101 \kappa + 74 \kappa^2) }{ 
  (1 + 6 \kappa) (1 + 11 \kappa) (2 + 13 \kappa) (1 + 17 \kappa) (3 + 17 \kappa) } P^\kappa_{0100000}
\\ &+&  \frac{ 144 (1 + 2 \kappa) (1 + 3 \kappa) (1 + 5 \kappa) (1 + 18 \kappa) }{ 
  (1 + 7 \kappa) (1 + 8 \kappa) (1 + 11 \kappa)^2 (2 + 17 \kappa) } P^\kappa_{0000001}
\\
P^\kappa_{1000000}\times P^\kappa_{0010000}&=&
 P^\kappa_{1010000}
+\frac{ 3 }{ 1 + 2 \kappa } P^\kappa_{0001000}
+\frac{ 16 (1 + 2 \kappa) (1 + 8 \kappa) }{ (1 + 5 \kappa) (1 + 7 \kappa) (2 + 7 \kappa) } P^\kappa_{1000010}
\\ &+&  \frac{ 15 (1 + \kappa) (1 + 3 \kappa) (1 + 11 \kappa) }{ (1 + 4 \kappa) (1 + 5 \kappa)^2 
   (1 + 7 \kappa) } P^\kappa_{0100001}
+  \frac{ 48 (1 + 2 \kappa) (1 + 4 \kappa) (2 + 17 \kappa) }{ 
  (1 + 7 \kappa) (1 + 8 \kappa) (1 + 9 \kappa) (3 + 16 \kappa) } P^\kappa_{2000000}
\\ &+&  \frac{ -24 (1 + 11 \kappa) (-5 - 187 \kappa - 2180 \kappa^2 - 9982 \kappa^3 - 18875 \kappa^4 - 
     11395 \kappa^5 + 1800 \kappa^6) }{ 
  (1 + \kappa) (1 + 5 \kappa)^3 (1 + 7 \kappa) (1 + 8 \kappa) (1 + 17 \kappa) (4 + 17 \kappa) } P^\kappa_{0010000}
\\ &+&  \frac{ 240 (1 + \kappa) (1 + 2 \kappa) (1 + 3 \kappa) (1 + 12 \kappa) (1 + 13 \kappa) }{ 
  (1 + 6 \kappa) (1 + 7 \kappa)^2 (1 + 8 \kappa) (2 + 11 \kappa) (3 + 17 \kappa) } P^\kappa_{0000010}
\\ &+&  \frac{ 384 (1 + 2 \kappa) (1 + 3 \kappa) (1 + 4 \kappa)^2 (1 + 5 \kappa) (1 + 12 \kappa) (1 + 14 \kappa) 
   (1 + 17 \kappa) }{ (1 + 7 \kappa)^2 (1 + 8 \kappa)^2 (1 + 9 \kappa) (1 + 11 \kappa) (2 + 11 \kappa) 
   (2 + 13 \kappa) (3 + 17 \kappa) } P^\kappa_{1000000}
\\
P^\kappa_{1000000}\times P^\kappa_{0000100}&=&
 P^\kappa_{1000100}
+\frac{ 5 }{ 1 + 4 \kappa } P^\kappa_{0100010}
+\frac{ 10 (1 + \kappa) (1 + 9 \kappa) }{ (1 + 4 \kappa)^2 (1 + 7 \kappa) } P^\kappa_{
  0010001}
\\ &+&  \frac{ 240 (1 + \kappa) (1 + 2 \kappa) (1 + 10 \kappa) }{ 
  (1 + 5 \kappa) (2 + 9 \kappa) (2 + 13 \kappa) (3 + 13 \kappa) } P^\kappa_{1100000}
+ \frac{ 32 (1 + 2 \kappa) (1 + 3 \kappa) (1 + 12 \kappa) }{ 
  (1 + 5 \kappa) (1 + 7 \kappa)^2 (2 + 11 \kappa) } P^\kappa_{0000011}
\\ &+&  \frac{ -90 (1 + 10 \kappa) (-16 - 562 \kappa - 5649 \kappa^2 - 18716 \kappa^3 - 18653 \kappa^4 + 
     3276 \kappa^5) }{ (1 + 4 \kappa) (1 + 7 \kappa) (2 + 9 \kappa) (2 + 13 \kappa) (3 + 13 \kappa) 
   (1 + 17 \kappa) (4 + 17 \kappa) } P^\kappa_{0000100}
\\ &+&  \frac{ 240 (1 + \kappa) (1 + 2 \kappa) (1 + 3 \kappa) (2 + 7 \kappa) (1 + 10 \kappa) (1 + 12 \kappa) 
   (2 + 17 \kappa) }{ (1 + 5 \kappa) (1 + 7 \kappa)^2 (1 + 8 \kappa) (2 + 9 \kappa) (2 + 11 \kappa) 
   (2 + 13 \kappa) (3 + 16 \kappa) } P^\kappa_{1000001}
\\ &+&  \frac{ 420 (1 + \kappa) (1 + 2 \kappa) (1 + 3 \kappa) (1 + 4 \kappa) (2 + 7 \kappa) (1 + 11 \kappa) 
   (1 + 12 \kappa) (1 + 14 \kappa) }{ (1 + 5 \kappa)^2 (1 + 6 \kappa) (1 + 7 \kappa) (1 + 8 \kappa) 
   (2 + 11 \kappa) (2 + 13 \kappa)^2 (3 + 17 \kappa) } P^\kappa_{0100000}
\\
P^\kappa_{1000000}\times P^\kappa_{0000010}&=&
 P^\kappa_{1000010}
+\frac{ 6 }{ 1 + 5 \kappa } P^\kappa_{0100001}
+\frac{ 27 (1 + 3 \kappa) }{ (1 + 8 \kappa) (1 + 11 \kappa) } P^\kappa_{000000
   2}
+\frac{ 15 (1 + \kappa) (1 + 11 \kappa) }{ (1 + 5 \kappa)^2 (1 + 9 \kappa) } P^\kappa_{
  0010000}
\\ &+&  \frac{ -48 (1 + 3 \kappa) (1 + 13 \kappa) (-2 - 45 \kappa - 109 \kappa^2 + 6 \kappa^3) }{ 
  (1 + \kappa) (1 + 6 \kappa) (1 + 7 \kappa) (1 + 9 \kappa) (1 + 17 \kappa) (3 + 17 \kappa) } P^\kappa_{
  0000010}
\\ &+&  \frac{ 120 (1 + \kappa) (1 + 3 \kappa) (1 + 4 \kappa) (1 + 6 \kappa) (1 + 14 \kappa) (1 + 17 \kappa) }{ 
  (1 + 5 \kappa) (1 + 7 \kappa) (1 + 8 \kappa) (1 + 9 \kappa)^2 (1 + 13 \kappa) (2 + 13 \kappa) } P^\kappa_{
  1000000}
\\
P^\kappa_{1000000}\times P^\kappa_{0000001}&=&
 P^\kappa_{1000001}
+\frac{ 7 }{ 1 + 6 \kappa } P^\kappa_{0100000}
+\frac{ 54 (1 + 3 \kappa) (1 + 18 \kappa) }{ (1 + 11 \kappa) (1 + 17 \kappa) (2 + 17 \kappa) } P^\kappa_{
  0000001}
\\
P^\kappa_{0100000}\times P^\kappa_{0100000}&=&
 P^\kappa_{0200000}
+\frac{ 2 }{ 1 + \kappa } P^\kappa_{0001000}
+\frac{ 8 (1 + 2 \kappa) }{ (1 + 3 \kappa) (1 + 5 \kappa) } P^\kappa_{100001
   0}
\\ &+&  \frac{ 12 (5 + 84 \kappa + 255 \kappa^2 + 160 \kappa^3) }{ 
  (1 + 5 \kappa)^2 (1 + 11 \kappa) (3 + 11 \kappa) } P^\kappa_{0100001}
+  \frac{ 32 (1 + 2 \kappa) (1 + 4 \kappa) }{ (1 + 5 \kappa) (1 + 7 \kappa) (1 + 9 \kappa) } P^\kappa_{
  2000000}
\\ &+&  \frac{ 144 (1 + 2 \kappa) (1 + 3 \kappa) (1 + 5 \kappa) (1 + 12 \kappa) }{ 
  (1 + 7 \kappa) (1 + 8 \kappa) (1 + 11 \kappa)^2 (2 + 11 \kappa) } P^\kappa_{000000
   2}
\\ &+&  \frac{ -40 (1 + 2 \kappa)^2 (-1 - 22 \kappa - 59 \kappa^2 + 10 \kappa^3) }{ 
  (1 + \kappa) (1 + 4 \kappa) (1 + 5 \kappa)^3 (1 + 11 \kappa) } P^\kappa_{0010000}
\\ &+&  \frac{ -160 (1 + 2 \kappa)^2 (1 + 3 \kappa) (1 + 12 \kappa) (-3 - 64 \kappa + 11 \kappa^2) }{ 
  (1 + \kappa) (1 + 6 \kappa) (1 + 7 \kappa) (1 + 11 \kappa)^2 (2 + 11 \kappa) (3 + 17 \kappa) } P^\kappa_{
  0000010}
\\ &+&  \frac{ 192 (1 + 2 \kappa) (1 + 3 \kappa) (1 + 4 \kappa) (1 + 14 \kappa) (5 + 101 \kappa + 74 \kappa^2) }{  (1 + 7 \kappa) (1 + 8 \kappa) (1 + 9 \kappa) (1 + 11 \kappa)^2 (2 + 13 \kappa) (3 + 17 \kappa) } P^\kappa_{
  1000000}
\\ &+&  \frac{ 1152 (1 + 2 \kappa) (1 + 3 \kappa) (1 + 4 \kappa) (1 + 5 \kappa) (1 + 6 \kappa) (1 + 18 \kappa) }{  (1 + 7 \kappa) (1 + 9 \kappa) (1 + 11 \kappa)^2 (1 + 13 \kappa) (1 + 17 \kappa) (2 + 17 \kappa) } 
\\
P^\kappa_{0100000}\times P^\kappa_{0000010}&=&
 P^\kappa_{0100010}
+\frac{ 5 }{ 1 + 4 \kappa } P^\kappa_{0010001}
+\frac{ 32 (1 + 2 \kappa) (1 + 12 \kappa) }{ (1 + 7 \kappa) (1 + 11 \kappa) (2 + 11 \kappa) } P^\kappa_{
  0000011}
\\ &+&  \frac{ 30 (1 + \kappa) (1 + 10 \kappa) }{ (1 + 5 \kappa) (1 + 9 \kappa) (2 + 9 \kappa) } P^\kappa_{
  1100000}
\\ &+&  \frac{ -60 (1 + 10 \kappa) (-3 - 56 \kappa - 119 \kappa^2 + 18 \kappa^3) }{ 
  (1 + 4 \kappa) (1 + 9 \kappa) (2 + 9 \kappa) (1 + 11 \kappa) (3 + 13 \kappa) } P^\kappa_{
  0000100}
\\ &+&  \frac{ 48 (1 + 2 \kappa) (1 + 3 \kappa) (14 + 531 \kappa + 6042 \kappa^2 + 23399 \kappa^3 + 
     24354 \kappa^4) }{ (1 + 5 \kappa) (1 + 7 \kappa) (1 + 9 \kappa) (2 + 9 \kappa) (1 + 11 \kappa) 
   (2 + 11 \kappa) (3 + 16 \kappa) } P^\kappa_{1000001}
\\ &+&  \frac{ -420 (1 + 2 \kappa) (1 + 3 \kappa) (1 + 4 \kappa) (1 + 12 \kappa) (1 + 14 \kappa) 
   (-3 - 64 \kappa + 11 \kappa^2) }{ (1 + 5 \kappa) (1 + 6 \kappa) (1 + 9 \kappa) (1 + 11 \kappa) 
   (2 + 11 \kappa) (1 + 13 \kappa) (2 + 13 \kappa) (3 + 17 \kappa) } P^\kappa_{010000
   0}
\\ &+&  \frac{ 864 (1 + \kappa) (1 + 2 \kappa) (1 + 3 \kappa) (1 + 4 \kappa) (1 + 14 \kappa) (1 + 18 \kappa) }{ 
  (1 + 7 \kappa) (1 + 8 \kappa) (1 + 9 \kappa) (1 + 11 \kappa)^2 (2 + 13 \kappa) (2 + 17 \kappa) } P^\kappa_{
  0000001}
\\
P^\kappa_{0100000}\times P^\kappa_{0000001}&=&
 P^\kappa_{0100001}
+\frac{ 6 }{ 1 + 5 \kappa } P^\kappa_{0010000}
+\frac{ 16 (1 + 2 \kappa) (1 + 13 \kappa) }{ (1 + 6 \kappa) (1 + 7 \kappa) (1 + 11 \kappa) } P^\kappa_{
  0000010}
\\&+& \frac{ 32 (1 + 2 \kappa) (1 + 4 \kappa) (1 + 17 \kappa) }{ 
  (1 + 7 \kappa) (1 + 8 \kappa) (1 + 9 \kappa) (1 + 11 \kappa) } P^\kappa_{1000000}
\\
P^\kappa_{0010000}\times P^\kappa_{0000001}&=&
 P^\kappa_{0010001}
+\frac{ 6 }{ 1 + 5 \kappa } P^\kappa_{1100000}
+\frac{ 20 (1 + \kappa) (1 + 10 \kappa) }{ (1 + 4 \kappa) (1 + 7 \kappa) (2 + 9 \kappa) } P^\kappa_{
  0000100}
\\ &+&  \frac{ 48 (1 + 2 \kappa) (1 + 3 \kappa) (1 + 12 \kappa) (2 + 17 \kappa) }{ 
  (1 + 7 \kappa)^2 (1 + 8 \kappa) (2 + 11 \kappa) (3 + 16 \kappa) } P^\kappa_{100000
   1}
\\ &+&  \frac{ 252 (1 + \kappa) (1 + 3 \kappa) (1 + 4 \kappa) (1 + 12 \kappa) (1 + 14 \kappa) }{ 
  (1 + 6 \kappa) (1 + 7 \kappa) (1 + 8 \kappa) (2 + 11 \kappa) (2 + 13 \kappa) (3 + 17 \kappa) } P^\kappa_{
  0100000}
\\
P^\kappa_{0000100}\times P^\kappa_{0000001}&=&
 P^\kappa_{0000101}
+\frac{ 4 }{ 1 + 3 \kappa } P^\kappa_{0001000}
+\frac{ 8 (1 + 2 \kappa) (1 + 9 \kappa) }{ (1 + 4 \kappa) (1 + 5 \kappa) (1 + 7 \kappa) } P^\kappa_{
  1000010}
\\ &+&  \frac{ 90 (1 + \kappa) (1 + 3 \kappa) (1 + 11 \kappa) }{ (1 + 5 \kappa)^2 (2 + 13 \kappa) 
   (3 + 13 \kappa) } P^\kappa_{0100001}
\\ &+&  \frac{ 40 (1 + \kappa) (1 + 2 \kappa) (2 + 7 \kappa) (1 + 10 \kappa) (1 + 11 \kappa) }{ 
  (1 + 5 \kappa)^3 (1 + 7 \kappa) (2 + 9 \kappa) (2 + 13 \kappa) } P^\kappa_{001000
   0}
\\ &+&  \frac{ 96 (1 + 2 \kappa) (1 + 3 \kappa) (1 + 4 \kappa) (2 + 7 \kappa) (1 + 9 \kappa) (1 + 12 \kappa) 
   (1 + 13 \kappa) }{ (1 + 5 \kappa) (1 + 6 \kappa) (1 + 7 \kappa)^2 (1 + 8 \kappa) (2 + 11 \kappa) 
   (2 + 13 \kappa) (3 + 17 \kappa) } P^\kappa_{0000010}
\\
P^\kappa_{0001000}\times P^\kappa_{0000001}&=&
P^\kappa_{0001001}
+\frac{ 5 }{ 1 + 4 \kappa } P^\kappa_{0110000}
+\frac{ 12 (1 + \kappa) (1 + 8 \kappa) }{ (1 + 3 \kappa) (1 + 5 \kappa) (2 + 7 \kappa) } P^\kappa_{
  1000100}
\\ &+&  \frac{ 60 (1 + \kappa) (1 + 2 \kappa) (1 + 6 \kappa) (1 + 8 \kappa) }{ 
  (1 + 4 \kappa) (1 + 5 \kappa)^2 (2 + 7 \kappa) (3 + 11 \kappa) } P^\kappa_{010001
   0}
\\ &+&  \frac{ 20 (1 + \kappa) (1 + 2 \kappa) (2 + 5 \kappa) (1 + 8 \kappa) (1 + 9 \kappa) (3 + 16 \kappa) }{ 
  (1 + 4 \kappa)^2 (1 + 5 \kappa)^3 (3 + 11 \kappa) (4 + 15 \kappa) } P^\kappa_{001000
   1}
\\ &+&  \frac{ 30 (1 + \kappa) (1 + 2 \kappa) (1 + 3 \kappa)^2 (1 + 8 \kappa) (1 + 10 \kappa) (3 + 17 \kappa) }{ 
  (1 + 4 \kappa)^2 (1 + 5 \kappa)^4 (2 + 9 \kappa) (3 + 13 \kappa) } P^\kappa_{110000
   0}
\\ &+&  \frac{ 40 (1 + \kappa) (1 + 2 \kappa) (1 + 3 \kappa) (2 + 5 \kappa) (1 + 8 \kappa) (1 + 9 \kappa) 
   (1 + 10 \kappa) (3 + 10 \kappa) (2 + 13 \kappa) }{ 
  (1 + 4 \kappa)^3 (1 + 5 \kappa)^3 (2 + 9 \kappa) (2 + 11 \kappa) (3 + 13 \kappa) (4 + 17 \kappa) } P^\kappa_{
  0000100}
  \\
  P^\kappa_{0000100}\times P^\kappa_{0000001}&=&
 P^\kappa_{0000101}
+\frac{ 4 }{ 1 + 3 \kappa } P^\kappa_{0001000}
+\frac{ 8 (1 + 2 \kappa) (1 + 9 \kappa) }{ (1 + 4 \kappa) (1 + 5 \kappa) (1 + 7 \kappa) } P^\kappa_{
  1000010}
\\ &+&\frac{ 90 (1 + \kappa) (1 + 3 \kappa) (1 + 11 \kappa) }{ (1 + 5 \kappa)^2 (2 + 13 \kappa) 
   (3 + 13 \kappa) } P^\kappa_{0100001}
\\ &+&  \frac{ 40 (1 + \kappa) (1 + 2 \kappa) (2 + 7 \kappa) (1 + 10 \kappa) (1 + 11 \kappa) }{ 
  (1 + 5 \kappa)^3 (1 + 7 \kappa) (2 + 9 \kappa) (2 + 13 \kappa) } P^\kappa_{001000
   0}
\\ &+&  \frac{ 96 (1 + 2 \kappa) (1 + 3 \kappa) (1 + 4 \kappa) (2 + 7 \kappa) (1 + 9 \kappa) (1 + 12 \kappa) 
   (1 + 13 \kappa) }{ (1 + 5 \kappa) (1 + 6 \kappa) (1 + 7 \kappa)^2 (1 + 8 \kappa) (2 + 11 \kappa) 
   (2 + 13 \kappa) (3 + 17 \kappa) } P^\kappa_{0000010}
\\
P^\kappa_{0000010}\times P^\kappa_{0000001}&=&
 P^\kappa_{0000011}
+\frac{ 3 }{ 1 + 2 \kappa } P^\kappa_{0000100}
+\frac{ 20 (1 + 3 \kappa) (1 + 10 \kappa) }{ (1 + 7 \kappa) (1 + 9 \kappa) (2 + 9 \kappa) } P^\kappa_{
  1000001}
\\ &+&  \frac{ 42 (1 + \kappa) (1 + 4 \kappa) (1 + 14 \kappa) }{ 
  (1 + 5 \kappa) (1 + 6 \kappa) (1 + 9 \kappa) (2 + 13 \kappa) } P^\kappa_{0100000}
\\ &+&  \frac{ 108 (1 + 3 \kappa) (1 + 4 \kappa) (1 + 6 \kappa) (1 + 14 \kappa) (1 + 18 \kappa) }{ 
  (1 + 8 \kappa) (1 + 9 \kappa) (1 + 11 \kappa) (1 + 13 \kappa) (2 + 13 \kappa) (2 + 17 \kappa) } P^\kappa_{
  0000001}
\\
P^\kappa_{0000001}\times P^\kappa_{0000001}&=&
 P^\kappa_{0000002}
+\frac{ 2 }{ 1 + \kappa } P^\kappa_{0000010}
+\frac{ 12 (1 + 4 \kappa) }{ (1 + 5 \kappa) (1 + 9 \kappa) } P^\kappa_{100000
   0}
\\&+&\frac{ 56 (1 + 4 \kappa) (1 + 8 \kappa) }{ (1 + 9 \kappa) (1 + 13 \kappa) (1 + 17 \kappa) } 
\\
\end{eqnarray*}
\section*{Acknowledgements} 
This work has been partially supported by the Spanish 
Ministerio de Educaci\'{o}n y Ciencia under grants BFM2003-02532 (J.F.N) and BFM2003-00936 / FISI (W.G.F and A.M.P). This paper was completed during the visit of A.M.P. to Max-Planck-Institut f\"ur Gravitationsphysik. He would like to thank the staff of the Institute for hospitality.

\end{document}